\documentclass[useAMS,usenatbib]{mn2e}
\usepackage{graphicx,txfonts,amssymb,amsfonts,natbib,times,hyperref}

\begin{document}

\title[Constraints on running]
{Primordial density perturbations with running spectral index: 
impact on non-linear cosmic structures}

\author[C.\,Fedeli et al.]{C.\,Fedeli$^{1,2,3}$, F.\,Finelli$^{4,3}$ and L.\,Moscardini$^{1,2,3}$\\$^1$ 
      Dipartimento di Astronomia, Universit\`a di Bologna,
      Via Ranzani 1, I-40127 Bologna, Italy (cosimo.fedeli@unibo.it)\\$^2$ INAF-Osservatorio
      Astronomico di Bologna, Via Ranzani 1, I-40127 Bologna, Italy\\$^3$
      INFN, Sezione di Bologna, Viale Berti Pichat 6/2, I-40127 Bologna, Italy\\$^4$
      INAF-IASF Bologna, Via Gobetti 101, I-40129 Bologna, Italy }
      
\maketitle

\begin{abstract}
We explore the statistical properties of non-linear cosmic structures in a flat $\Lambda$CDM cosmology in which the index $n_\mathrm{S}$ of the primordial power spectrum for scalar perturbations is allowed to depend on the scale. Within the inflationary paradigm, the running  of the scalar spectral index can be related to the properties of the inflaton potential, and it is hence of critical importance to test it with all kinds of observations, which cover the linear and non-linear regime of gravitational instability. We focus on the amount of running $\alpha_{\mathrm{S},0}$ ($\equiv d n_\mathrm{S}/d\ln (k/k_\mathrm{p})$) allowed by an updated combination of CMB anisotropy data and the 2dF Galaxy Redshift Survey. Our analysis constrains $\alpha_{\mathrm{S},0} = -0.051^{+0.047}_{-0.053}$ $\left(-0.034^{+0.039}_{-0.040}\right)$ at $95 \%$ Confidence Level when (not) taking into account primordial gravitational waves in a ratio as predicted by canonical single field inflation, in agreement with other works. For the cosmological models best fitting the data both with and without running we studied the abundance of galaxy clusters and of rare objects, the halo bias, the concentration of dark matter halos, the Baryon Acoustic Oscillation, the power spectrum of cosmic shear, and the Integrated Sachs-Wolfe effect. We find that counting galaxy clusters in future X-ray and Sunyaev-Zel'dovich surveys could discriminate between the two models, more so if broad redshift information about the cluster samples will be available. Likewise, measurements of the power spectrum of cosmological weak lensing as performed by planned all-sky optical surveys such as EUCLID could detect a running of the primordial spectral index, provided the uncertainties about the source redshift distribution and the underlying matter power spectrum are well under control.
\end{abstract}

\section{Introduction}

The formation of cosmic structures occurred via gravitational instability. 
Small density fluctuations in the dark matter fluid grew up in time, and eventually detached 
from the overall expansion of the Universe, collapsing and giving rise to dark matter halos. 
This kind of process is relatively well understood, thanks to both perturbation theory in the 
linear and quasi-linear regimes, and numerical simulations in the fully non-linear regime. 
The additional role of baryonic matter, that fell into newly formed dark matter potential 
wells is less well understood, but important in order to reliably extract cosmological information from future cosmological surveys (\citealt{JI06.1,LI06.1}; \citealt*{RU08.2,ST09.1}).

The evolution with redshift of cosmic structures is mainly determined, in a statistical sense, by the initial 
conditions and by the subsequent expansion history of the Universe. The latter depends critically on the matter and energy content of the Universe itself, and this dependence is particularly important in order to gauge the dynamical evolution of dark energy  \citep*{CU09.1,GR09.2,SA09.1}. Concerning the initial conditions, the simplest models of inflation predict that the primordial density fluctuations should be distributed according to a nearly Gaussian statistics, with a nearly Harrison-Zeldovich (i.e., scale-invariant) spectrum. Cosmic Microwave Background radiation (CMB) and Large Scale Structure (LSS) observations are continuously improving the measurement of deviations from this Gaussian scale invariant case. Presently deviations from Gaussian statistics of the CMB pattern are not observed \citep{KO10.1}, whereas the Harrison-Zeldovich spectrum is disfavoured with a statistical 
significance greater than $99.9 \%$ Confidence Level (CL) from a recent compilation of CMB and LSS data \citep{FI09.1}. The evolution of non-linear structures depends crucially on the spectral index of scalar perturbations and their distribution, and recently much attention has been payed to deviations from Gaussianity as these can put constraints on the inflationary model (e.g., \citealt*{FE09.1}; \citealt{GR09.1,VE09.1}). 

Measures of deviations from an exact power-law for the spectrum of scalar perturbations have been searched in CMB and LSS with different techniques. The simplest parametrization of this deviation is a logarithmic dependence on the wavenumber of the spectral index $n_\mathrm{S}$ \citep{KO95.1}, i.e. $n_\mathrm{S} (k) = n_\mathrm{S} (k_\mathrm{p}) + \alpha_{\mathrm{S},0} \ln (k/k_\mathrm{p})$, where $\alpha_{\mathrm{S},0}$ is called {\em running of the scalar spectral index} and $k_\mathrm{p}$ 
is a pivot scale. Inflation gives an important support to the wavelength dependence of the spectral indices of scalar and tensor perturbations, which would mainly depend on third and higher-order derivatives of the inflaton potential, whereas the magnitudes of the spectral indices themselves depend mainly on its derivatives up to second order \citep{LI00.1}. The simplest canonical single field slow-roll models predict the running of the spectral indices to be of the order of a few $\times\; 10^{-4}$, whereas current CMB and LSS data are sensitive only at values for the scalar running of the order of a few $\times\;10^{-2}$ \citep{PA06.1,PA07.2,FI09.1,KO09.1,KO10.1}. 

CMB and LSS present data do not require the introduction of the running to improve drastically the $\chi^2$, but substantial values at $95 \%$ CL for the running are allowed and also weakly favoured by different and indipendent analyses of current data (\citealt*{FI06.1,FE07.3,LE07.1}; \citealt{KI09.1,FI09.1}). These allowed values for the running might be just due to the degenaracy in cosmological parameters \citep{EF99.1}, but at the time being cosmological scenarios with and without running provide a very similar explanation of current observations. It is hence important to figure out probes complementary to the combination of CMB and LSS that might be useful to discriminate between these two scenarios (see \citet{IS04.1} for an earlier work studying this issue in the context of weak lensing). This is the subject of the present work, where we put particular emphasis on the non-linear, large scale matter distribution that results from a running in the primordial power spectrum of scalar perturbations. In practically addressing how the running of the spectral index affects non-linear observables, we computed the latter in two cosmological models which are best-fits of CMB and 2dFGRS data with zero and non-zero running, respectively. This analysis goes beyond the simple choice of a cosmological model with the relative variation of the running of the spectral index keeping all the other cosmological parameters fixed and it suffices for selecting which non-linear probes are more promising in constraining a running spectral index for values in the ballpark allowed by current CMB and LSS data. Note that the quantitative forecast of a future probe of the non-linear regime on the measurement of the running of the scalar spectral index is beyond the the scope of the present paper.

The rest of the paper is organized as follows. In Section \ref{sct:cosmoparams} we describe in details the two cosmological models considered in this work, namely a standard $\Lambda$CDM model and a model with running spectral index. We have obtained the parameter sets for these models by a combined analysis of various CMB datasets with LSS input. In Section \ref{sct:observables} we detail the cosmological observables that have been addressed here, and what kind of discriminating power are they expected to give between the two models. In Section \ref{sct:results} we present our results, that are discussed in Section \ref{sct:alternative}. Finally, in Section \ref{sct:conclusions} we summarize our conclusions. We shall adopt the natural units system, where $c = 1$, unless otherwise stated.

\section{Cosmological Parameters}\label{sct:cosmoparams}

As mentioned in the introduction, we estimated the values of cosmological parameters by combining different CMB datasets with LSS data from the 2dFGRS \citep{CO05.1}. We have chosen to use the WMAP-5 year \citep{DU09.1,NO09.1}, ACBAR \citep{RE09.1}, BICEP \citep{CH10.1} and QUaD \citep{BR09.1} CMB anisotropy data. In order to avoid  correlations between different datasets which cover the same region of the sky, we removed in the analysis the following CMB band powers: a) all the QUaD TT band powers since they overlap with data from the `CMB8' region of ACBAR; b) the ACBAR band powers with $\ell<790$ and $\ell>1950$ to avoid overlap with WMAP (which is cosmic variance limited up to $\ell=530$, see \citealt{NO09.1}) and contamination from foreground residuals, 
respectively; c) the QUaD TE band powers which overlap with WMAP ones and the QUaD EE band powers which overlap with BICEP; d) the BICEP TT and TE band powers (i.e., we used just EE and BB information from BICEP). Due to the dependence of the matter power spectrum shape on $n_\mathrm{S}$, $\Omega_{\mathrm{m},0}$, and $\Omega_{\mathrm{b},0}$, the addition of 2dFGRS data \citep{CO05.1} to our CMB compilation helps in decreasing the degeneracies among the cosmological parameters we vary in our analysis.

We used the \texttt{CosmoMC}\footnote{\sc http://cosmologist.info/cosmomc/} code (\citealt*{LE00.1,LE02.1}) in order to compute the Bayesian probability distribution of cosmological parameters. We varied the baryon density parameter $\omega_\mathrm{b}\equiv\Omega_{\mathrm{b},0} h^2$, the cold dark matter density parameter $\omega_\mathrm{c}= \Omega_{\mathrm{c},0}h^2$ 
(with $h$ being $H_0/100$ km s$^{-1}$ Mpc$^{-1}$ and $\Omega_{\mathrm{m},0}=\Omega_{\mathrm{c},0}+\Omega_{\mathrm{b},0}$ being the total matter density parameter), the reionisation optical depth $\tau$, the ratio of the sound horizon to the angular dimater distance at 
decoupling $\theta$, $\ln ( 10^{10} A_\mathrm{S})$, $n_\mathrm{S}$ and $\alpha_{\mathrm{S},0}$, the latter being the parameters characterizing the power spectrum of curvature perturbations ${\mathcal R}$,

\begin{equation}\label{eqn:spectrum}
\frac{k^3}{2\pi^2}P_{\cal R} (k) = A_\mathrm{S} \left( \frac{k}{k_\mathrm{p}} \right)^{n_\mathrm{S}-1+\alpha_\mathrm{S}(k)}.
\end{equation}
In Eq. (\ref{eqn:spectrum}), $\alpha_\mathrm{S}(k) = \left(\alpha_{\mathrm{S},0}/2\right) \ln \left( k/k_\mathrm{p} \right)$, $A_\mathrm{S}$ is the amplitude of scalar perturbations at a given pivot scale $k_\mathrm{p}$, $n_\mathrm{S} \equiv n_\mathrm{S}(k_\mathrm{p})$ is the spectral index and $\alpha_{\mathrm{S},0}$ is its running. When included, we assumed the following spectrum for gravitational waves:

\begin{equation}
\frac{k^3}{2\pi^2}P_\mathrm{T} (k) = r A_\mathrm{S} \left( \frac{k}{k_\mathrm{p}} \right)^{n_\mathrm{T}+\alpha_\mathrm{T}(k)},
\end{equation}
where $\alpha_\mathrm{T}(k)\equiv \left(\alpha_{\mathrm{T},0}/2\right) \ln \left( k/k_\mathrm{p} \right)$ and $r$ is the tensor-to-scalar ratio. The tensor spectral index has been fixed through the first-order consistency condition $n_{\rm T} = -r/8$ when power-law spectra are assumed or through the second-order consistency condition

\begin{equation}\label{eqn:cond1}
n_{\rm T} = -\frac{r}{8} \left( 2 - \frac{r}{8} - n_{\rm S} \right)
\end{equation}
when running is included. The running of the tensor spectral index has been fixed to

\begin{equation}\label{eqn:cond2}
\alpha_{\mathrm{T},0} = \frac{r}{8} \left( \frac{r}{8} + n_{\rm S}-1 \right).
\end{equation}

We assumed a flat universe, a CMB temperature $T_{\rm CMB}=2.725$ K, three neutrino species with a 
negligible mass and we set the primordial Helium fraction to $y_{\rm He}=0.24$. The pivot scale of the primordial scalar and tensor power spectra was set to $k_\mathrm{p}=0.018$~Mpc$^{-1}$, nearly as recommended by \citet*{CO07.1}, and we have verified that this choice of pivot scale is still close to optimum for the combination of data used in the present analysis. We also assumed a top-hat prior on the age of the Universe, $10 \:{\rm Gyrs} \le t_0 \le 20\: {\rm Gyrs}$, and a Gaussian prior 
on the present Hubble parameter $H_0= 72 \pm 8$ Km s$^{-1}$ Mpc$^{-1}$ \citep{FR00.1}.

We used the lensed CMB and matter power spectra and we followed the method implemented in \texttt{CosmoMC} consisting in varying a nuisance parameter $A_{\rm SZ}$ which accounts for the
unknown amplitude of the thermal SZ \citep{SU72.1} contribution to the small-scale CMB data points assuming the model of \cite{KO02.1}.  We used a \texttt{CAMB}\footnote{\sc http://camb.info/} accuracy setting of $1.8$ and sampled the posterior using the Metropolis-Hastings algorithm \citep{HA70.1} at a temperature $T=2$ (for improved exploration of the tails), generating eight parallel chains and imposing a conservative Gelman-Rubin convergence criterion \citep{GE92.1} of $R-1 < 0.02$.

\begin{table}
\centering
\caption{\label{tab:margnor}
Mean parameter values and bounds of the central 68\%-credible intervals for the $\Lambda$CDM model and the model with running spectral index in absence of gravitational waves. For the running of the spectral index $\alpha_{\mathrm{S},0}$ the bounds of the central 95\%-credible intervals are quoted.}
\begin{tabular}{|l|cc|}
\hline
\hline
& $\Lambda$CDM ($\alpha_{\mathrm{S},0}=0$)&running\\
\hline
\hline
$\omega_{\rm b}$ & $0.0225^{+0.0006}_{-0.0005}$ & $ 0.0219^{+0.0006}_{-0.0005}$ \\
$\omega_{\rm c}$ & $0.109 \pm 0.004$    &   $0.113 \pm 0.005$  \\
$\tau$ & $0.087^{+0.007}_{-0.008}$  &   $0.094^{+0.008}_{-0.009}$ \\
$\ln \left[ 10^{10} A_{\rm S}\right]$ & $3.10^{+0.03}_{-0.04}$
& $3.14^{+0.05}_{-0.04}$ \\
$n_{\rm S}$ &$0.958 \pm 0.013$ & $0.952^{+0.014}_{-0.013}$  \\
$\alpha_{\mathrm{S},0}$ &--& $-0.034^{+0.039}_{-0.040}$  \\
\hline
\hline
$h$ &  $0.722 \pm 0.018$ &  $0.701^{+0.022}_{-0.021}$   \\
$\Omega_{\mathrm{m},0}$ & $0.253^{+0.019}_{-0.020}$ & $0.275 \pm 0.025$ \\       
$\sigma_8$ & $0.784 \pm 0.026$  & $0.792^{+0.025}_{-0.024}$\\
\hline
\hline
\end{tabular}
\end{table}

\begin{table}
\centering
\caption{\label{tab:nor}
Best-fit parameter values and standard deviations for the $\Lambda$CDM model and running model in absence of gravitational waves.}
\begin{tabular}{|l|cc|}
\hline
\hline
& $\Lambda$CDM ($\alpha_{\mathrm{S},0}=0$)&running\\
\hline
\hline
$\omega_{\rm b}$ & $0.0226 \pm 0.0005$ & $ 0.0222 \pm 0.0006$ \\   
$\omega_{\rm c}$ & $0.109 \pm 0.004$    &   $0.112 \pm 0.005$  \\
$\tau$ & $0.083 \pm 0.016$  &   $0.087 \pm 0.018$ \\
$\ln \left[ 10^{10} A_{\rm S}\right]$ & $3.09 \pm 0.03$  
& $3.12 \pm 0.05$ \\
$n_{\rm S}$ &$0.958 \pm 0.013$ & $0.953 \pm 0.013$  \\
$\alpha_{\mathrm{S},0}$ &--& $-0.023 \pm 0.020$  \\
\hline
\hline
$h$ &  $0.724 \pm 0.018$ &  $0.709 \pm 0.022$   \\
$\Omega_{\mathrm{m},0}$ & $0.251 \pm 0.020$ & $0.267 \pm 0.026$ \\ 
$\sigma_8$ & $0.783 \pm 0.026$  & $0.789 \pm 0.025$\\        
\hline
\hline
\end{tabular}
\end{table}

\begin{table}
\centering   
\caption{\label{tab:margr}
Mean parameter values and bounds of the central 68\%-credible intervals for the $\Lambda$CDM model and the model with running spectral index including gravitational waves. For the running of the spectral index $\alpha_{\mathrm{S},0}$ and the tensor-to-scalar ratio $r$ the bounds of the central 95\%-credible intervals are quoted.}
\begin{tabular}{|l|cc|}
\hline
\hline
& $\Lambda$CDM ($\alpha_{\mathrm{S},0}=0$)&running\\
\hline
\hline
$\omega_{\rm b}$ & $0.0228^{+0.006}_{-0.005}$ & $0.0223^{+0.006}_{-0.007}$ \\
$\omega_{\rm c}$ & $0.107 \pm 0.004$ & $0.112 \pm 0.005$  \\
$\tau$ & $0.088^{+0.007}_{-0.009}$  &   $0.010 \pm 0.001$ \\
$\ln \left[ 10^{10} A_{\rm S}\right]$ & $3.08^{+0.07}_{-0.009}$ & $3.15 \pm 0.05$ \\
$n_{\rm S}$ &$0.969 \pm 0.014$& $0.970^{+0.019}_{-0.018}$  \\
$r$ & $< 0.27$ & $ < 0.37$ \\
$\alpha_{\mathrm{S},0}$ &--& $-0.051^{+0.047}_{-0.053}$  \\
\hline
\hline
$h$ &  $0.717 \pm 0.020$ &  $0.709 \pm 0.023$   \\
$\Omega_{\mathrm{m},0}$ & $0.243 \pm 0.019$ & $0.269 \pm 0.026$ \\
$\sigma_8$ & $0.780^{+0.026}_{-0.025}$  & $0.793 \pm 0.028$ \\
\hline
\hline
\end{tabular}
\end{table}

\begin{figure}
        \includegraphics[width=\hsize]{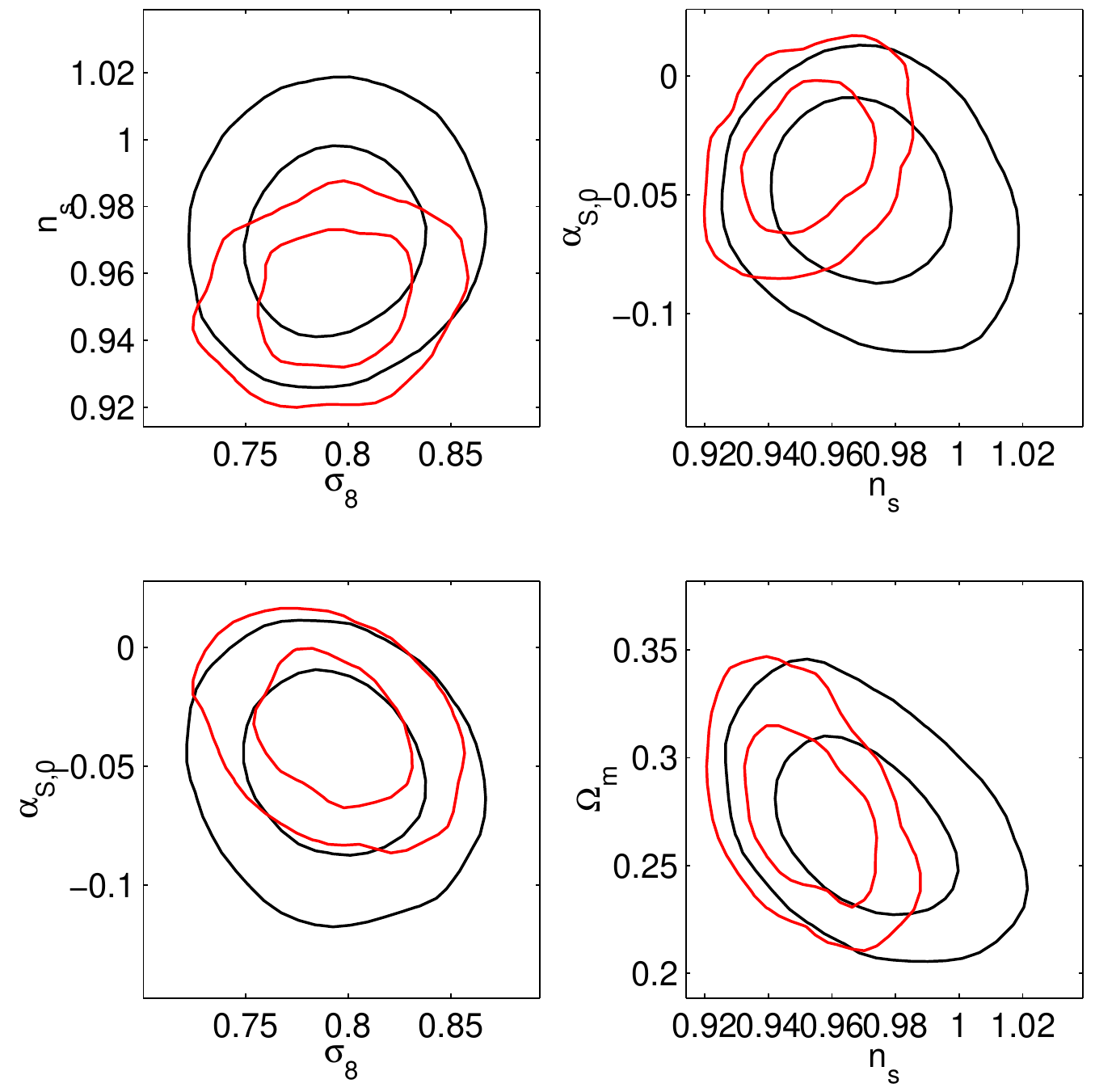}\hfill
        \caption{Marginalized constraints at $68 \%$ and $95 \%$ CL in the planes $(\sigma_8,n_\mathrm{S})$ (upper left panel), $(n_\mathrm{S},\alpha_{\mathrm{S},0})$ (upper right), $(\sigma_8,\alpha_{\mathrm{S},0})$ (lower left) and $(n_\mathrm{S},\Omega_{\mathrm{m},0})$ (lower right), for the case with (black lines) and without (red lines) gravitational waves.}
        \label{fig:2dplots}
\end{figure}

\begin{table}
\centering
\caption{\label{tab:r}
Best-fit parameter values and standard deviations for the $\Lambda$CDM model and running model including gravitational waves.}
\begin{tabular}{|l|cc|}
\hline
\hline
& $\Lambda$CDM ($\alpha_{\mathrm{S},0}=0$) &running\\
\hline
\hline
$\omega_{\rm b}$ & $0.0226 \pm 0.0006$ & $0.0221 \pm 0.0007$ \\
$\omega_{\rm c}$ & $0.110 \pm 0.004$    &   $0.112 \pm 0.005$  \\
$\tau$ & $0.084 \pm 0.017$  &   $0.093 \pm 0.020$ \\
$\ln \left[ 10^{10} A_{\rm S}\right]$ & $3.05 \pm 0.04$ & $3.13 \pm 0.05$ \\
$n_{\rm S}$ &$0.961 \pm 0.015$& $0.954 \pm 0.018$  \\
$r$ & $0.014 \pm 0.07$ & $0.022 \pm 0.11$ \\ 
$\alpha_{\mathrm{S},0}$ &--& $-0.032 \pm 0.025$  \\
\hline
\hline
$h$ &  $0.717 \pm 0.020$ &  $0.705 \pm 0.024$   \\
$\Omega_{\mathrm{m},0}$ & $0.258 \pm 0.019$ & $0.270 \pm 0.027$ \\
$\sigma_8$ & $0.790 \pm 0.026$  & $0.788 \pm 0.028$ \\
\hline
\hline
\end{tabular}
\end{table}

Our results for the mean values and the best-fit values of the cosmological parameters are presented in Tables \ref{tab:margnor} and \ref{tab:nor} (\ref{tab:margr} and \ref{tab:r}) for the case without (with) gravitational waves. In each Table we report the parameter sets for two cosmologies, the standard $\Lambda$CDM and the model with the running. These results are consistent with those obtained in \citet{FI09.1} by using the SDSS-7 LRG halo correlation function computed in \citet{RE10.1}; \citet*{RE09.3} instead of the 2dFGRS data \citep{CO05.1} and the distinctions are due to the long studied differences between the two galaxy surveys \citep{PE07.1}. In the case without (with) gravitational waves the inclusion of a wavelength dependence of the spectral indices as predicted by canonical single field inflation is not required and we found an improvement of $\Delta \chi^2 \sim 2 \, (\sim4)$. In Figure \ref{fig:2dplots} we report the marginalized constraints derived by our analysis concerning the parameters $\sigma_8$, $\Omega_{\mathrm{m},0}$, $n_\mathrm{S}$ and $\alpha_{\mathrm{S},0}$.

\begin{figure*}
        \includegraphics[width=0.45\hsize]{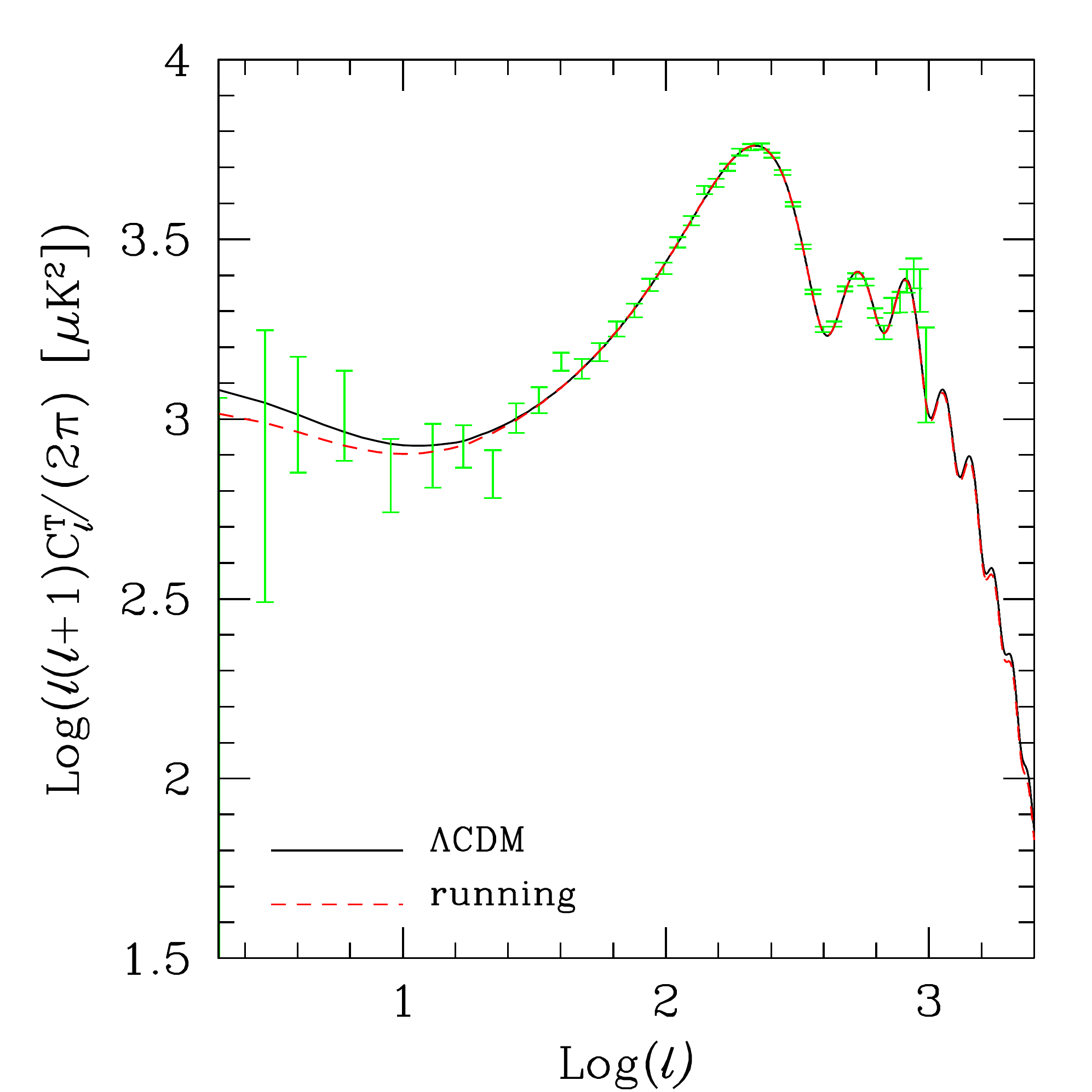}\hfill
        \includegraphics[width=0.45\hsize]{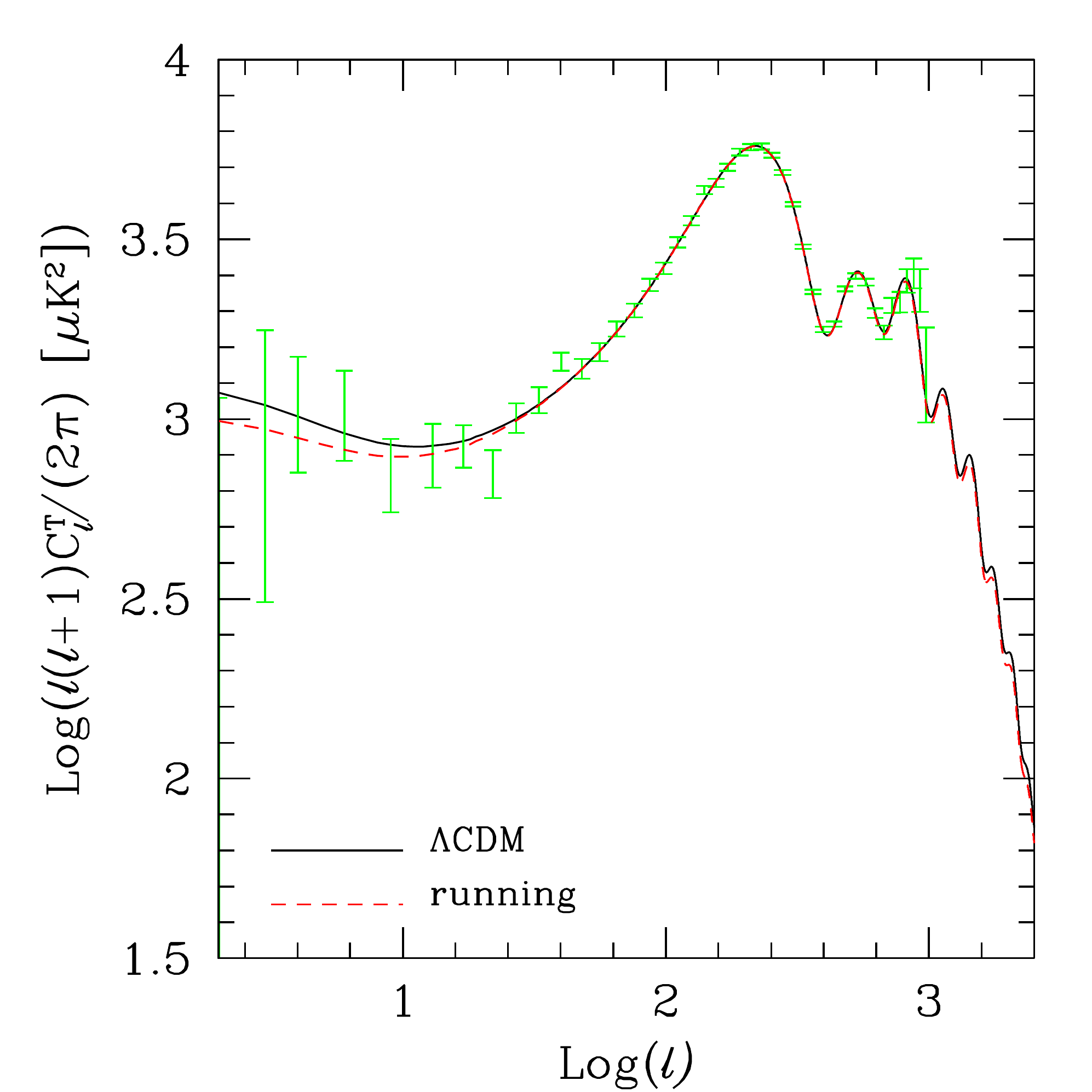}
        \caption{\emph{Left panel}. The CMB temperature anisotropies best-fit power spectra 
in absence of gravitational waves. The black solid (red dashed) line corresponds to the cosmological model without (with) running of the primordial spectral index, as labeled. \emph{Right panel}. The same as the left panel but taking into account gravitational waves. In both panels, the green bars represent the errors on the binned power spectrum measured by WMAP \citep{KO09.1}.}
        \label{fig:Cls}
\end{figure*}

These results show that the central value of the spectral index $n_\mathrm{S}$ almost does not vary when running is included, in contrast to large shifts towards blue values obtained, e.g., in the WMAP-5 year analysis (see the discussion in Section \ref{sct:alternative}). This is a benefit of choosing an optimal pivot scale for the analysis, which removes most of the non-necessary degenaracies in the $(n_\mathrm{S}, \alpha_{\mathrm{S},0})$ plane.

In the remainder of this paper we adopted as our fiducial pair of models those summarized in Table \ref{tab:nor}, namely the two sets of cosmological parameters that are returned as best fit by our analysis in absence of tensor modes. As the reader will notice, this model has the smallest amount of running $\alpha_{\mathrm{S},0}$ among those presented here. Our purpose is to be conservative in this respect, while in Section \ref{sct:alternative} we shall discuss how our results compare with different choices of cosmological model pairs. Given this preference, we will refer to our fiducial model without running primordial spectral index alternatively as "standard model", "concordance model" or "$\Lambda$CDM model".

In order to show the degeneracy of the models output of our analysis with respect to CMB observables, in Figure \ref{fig:Cls} we report the CMB temperature power spectra for the $\Lambda$CDM cosmology and the model with running spectral index, both in the case of absence or presence of primordial gravitational waves. The green bars represent the errors on the binned data points that have been measured by the WMAP satellite. It appears evident that the two CMB spectra are almost identical, apart from the very large scale part, where however the cosmic variance erases all the useful information. 

\begin{figure*}
        \includegraphics[width=0.45\hsize]{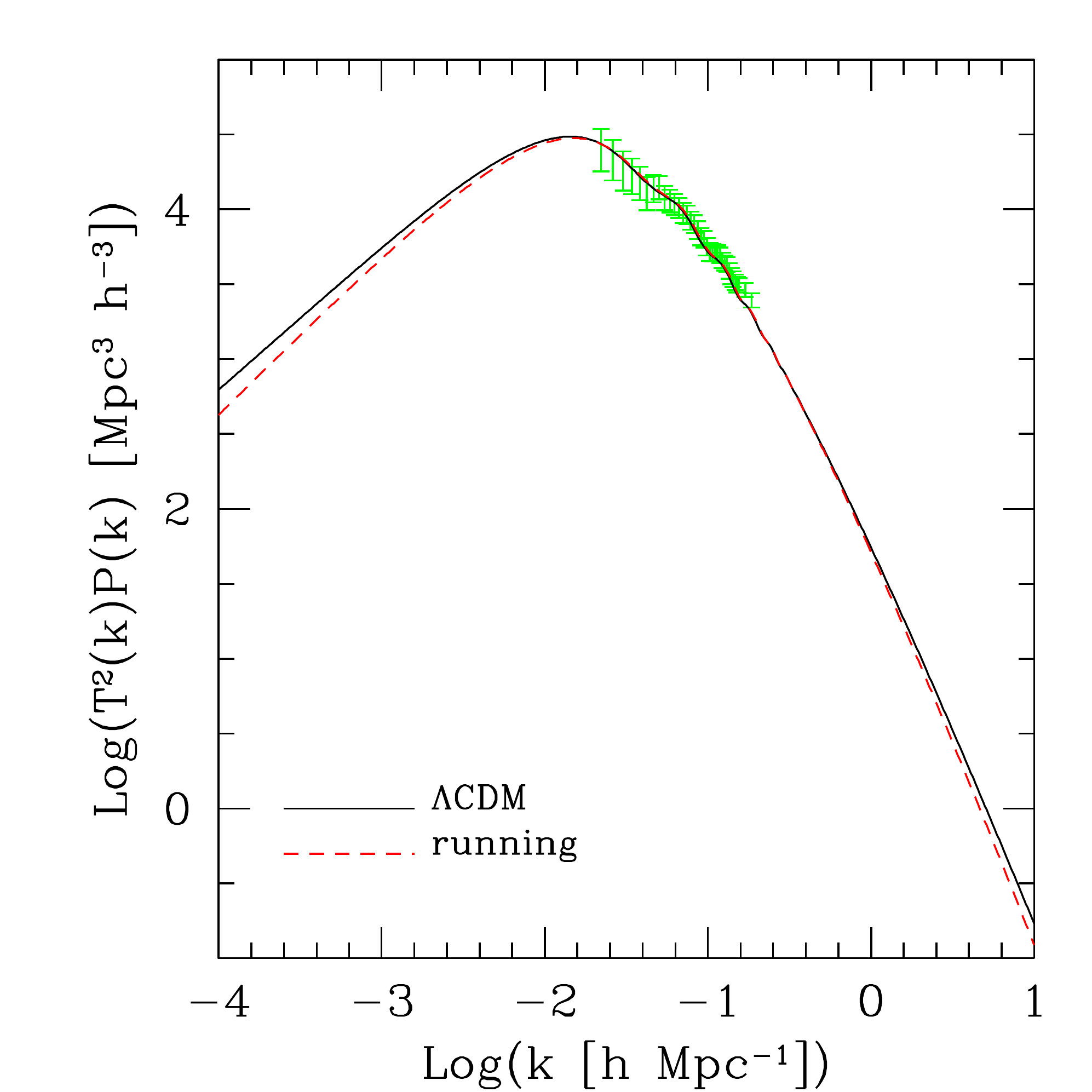}\hfill
        \includegraphics[width=0.45\hsize]{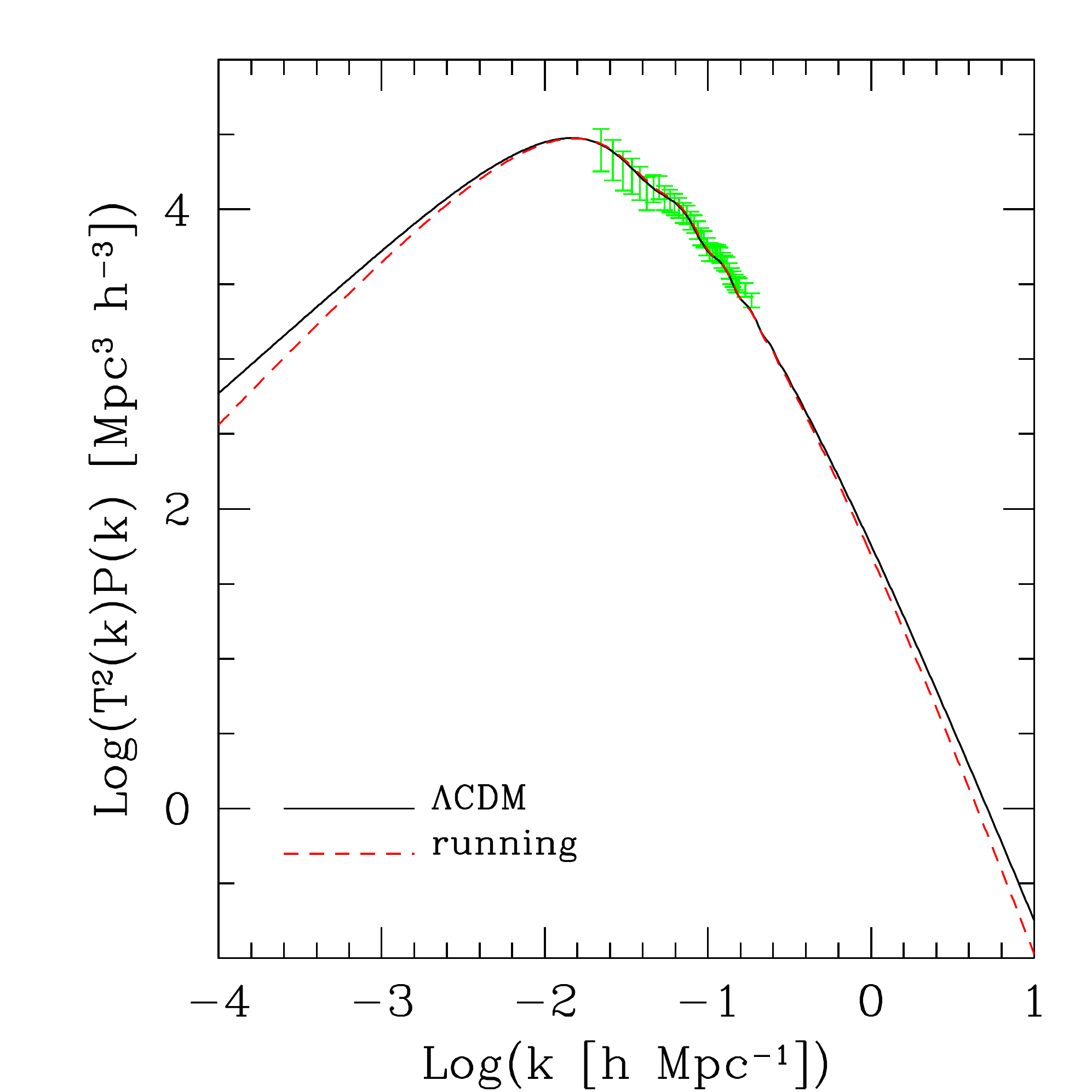}
        \caption{\emph{Left panel}. The linear cold dark matter best-fit power spectra at $z=0$  
in absence of gravitational waves. The black solid (red dashed) line corresponds to the cosmological model without (with) running of the primordial spectral index, as labeled. \emph{Right panel}. The same as the left panel but taking into accoung gravitational waves. In both panels the green bars represent the errors on the 2dFGRS galaxy power spectrum, scaled according to the bias factor $b = 1.26$ \citep{CO05.1}.}
        \label{fig:matter}
\end{figure*}

Further, in Figure \ref{fig:matter} we show the linear matter power spectra for the same two pairs of cosmological models (see Section \ref{sct:observables}). As can be seen, it is generically found that the models with running of the primordial spectral index have less power on very large and very small scales compared to standard cosmologies. This kind of trend has consequences on the process of structure formation that will be illustrated below.  

\section{Observables}\label{sct:observables}

In this Section we describe the cosmological observables that have been considered in this work. The results will be deferred to the next Section \ref{sct:results}. One aspect of the calculation of these observables that requires particular care is the relation between the power spectrum of curvature perturbations, that is inferred right away by our statistical analysis, and that of matter density fluctuations, that is instead more directly related to cosmic structure formation. The running of primordial spectral index adds further complication to this relation.

\subsection{Galaxy cluster abundance}\label{sct:obs:abundance}

One of the most obvious probes of the linear and non-linear stages of structure formation is the counting of galaxy clusters. The mass function of massive objects depends exponentially on the variance of the primordial power spectrum, hence a variation thereof is expected to have an impact of some kind.

The mass function $n(M,z)$ is the number of structures within the unit mass around $M$ that at redshift $z$ is contained in the unit comoving volume. An often used prescription for the mass function is the one of \cite{PR74.1}, that however tends to overpredict halo abundance at low masses with respect to the results of numerical $n$-body simulations. Other prescriptions exist, whose parameters are fitted against $n$-body simulations \citep{JE01.1,WA06.1,TI08.1} or determined based on more realistic models for the collapse of density perturbations (\citealt*{SH01.1,SH02.1}).  In the remainder of this work we adopted the latter prescription, according to which the mass function can be written as

\begin{eqnarray}\label{eqn:mf}
n(M,z) &=& A\sqrt{a} \frac{\rho_\mathrm{m}}{M} \frac{\delta_\mathrm{c}(z)}{\sqrt{2\pi}\sigma_M^3} \left[ 1 + \left( \frac{\sigma_M^2}{a\delta_\mathrm{c}(z)}\right)^{2p} \right] \times
\nonumber\\
&\times& \left| \frac{d\sigma_M^2}{dM} \right| \exp\left(-\frac{a\delta_\mathrm{c}^2(z)}{2\sigma^2_M}\right).
\end{eqnarray}
In Eq. (\ref{eqn:mf}) $\delta_\mathrm{c}(z) = \Delta_\mathrm{c}/D_+(z)$, where the quantity $\Delta_\mathrm{c}$ is the linear density threshold for spherical collapse, that is constant in an Einstein-de Sitter cosmology and only mildly dependent on redshift in models with a cosmological constant. We included this redshift dependence in our calculations but we do not indicate it explicitely, since it is practically irrelevant. The function $D_+(z)$ is the linear growth factor for density fluctuations, $\sigma_M$ is the \emph{rms} of density fluctuations smoothed on a scale corresponding to mass $M$, and $\rho_\mathrm{m}$ is the mean comoving matter density. The original \cite{PR74.1} formulation is obtained by setting $A = 0.5$, $a = 1$ and $p = 0$ in Eq. (\ref{eqn:mf}), while the \cite{SH02.1} revision is obtained with the values $A = 0.32$, $a = 0.75$ and $p = 0.3$.

\begin{figure}
	\includegraphics[width=\hsize]{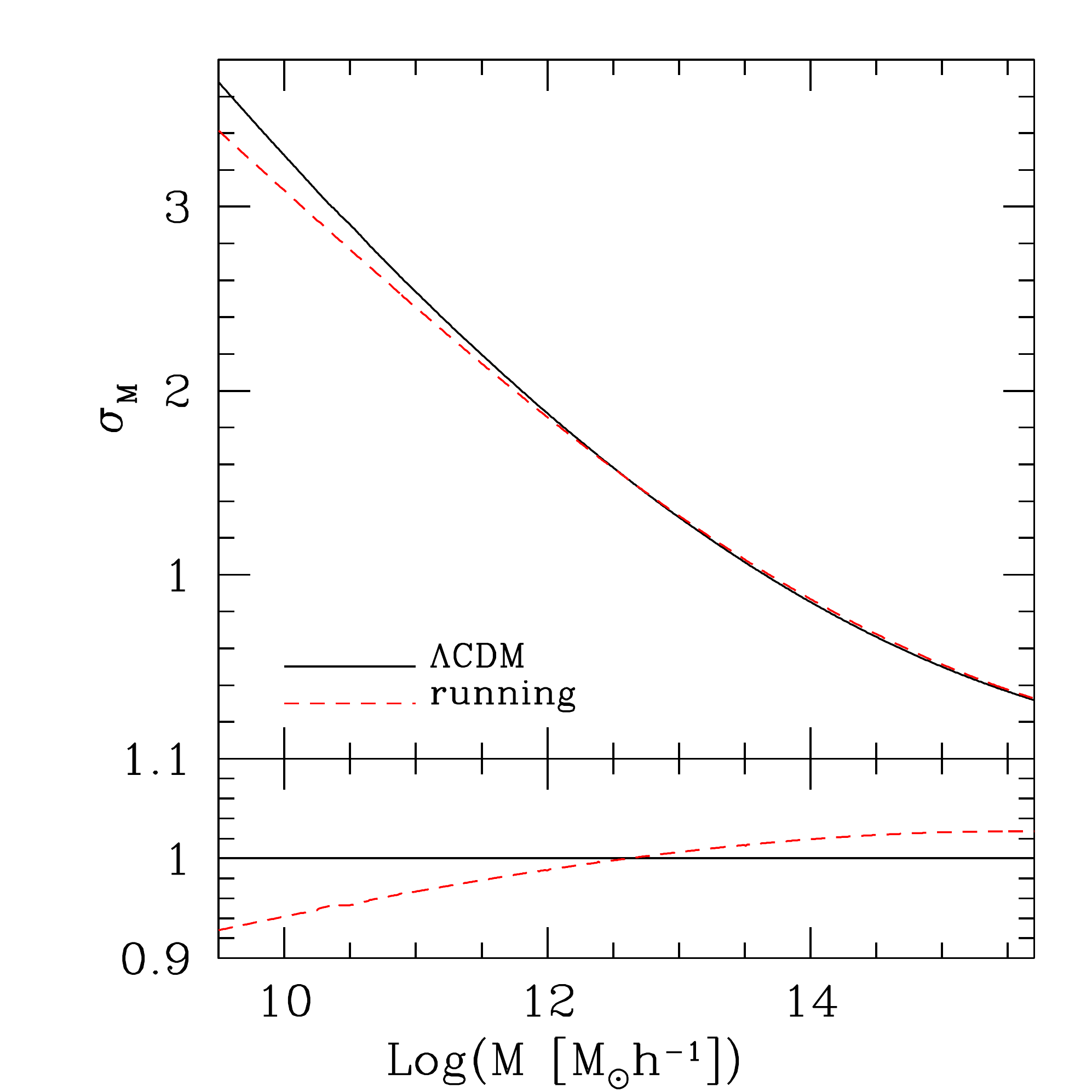}
	\caption{The \emph{rms} of matter density fluctuations as a function of the mass scale over which the density fluctuations are smoothed upon. Results for the two fiducial cosmological models are shown, as labeled in the plot. The lower inset shows the ratio between the two curves in the main panel.}
	\label{fig:rmsMass}
\end{figure}

In Figure \ref{fig:rmsMass} we show how the \emph{rms} of density fluctuations depends on the mass scale for the two cosmological models under consideration. This dependency reads

\begin{equation}\label{eqn:variance}
\sigma_M^2 \equiv \frac{1}{2\pi^2} \int_0^{+\infty} k^2dk W_R^2(k)T^2(k) P(k),
\end{equation}
where $T(k)$ is the matter transfer function. For that we adopted the semi-analytic fit given by \citet{EI98.1}, that accounts for all kinds of baryonic effects. A simplified expression, that ignores such effects, can be written as (see also \citealt{BA86.1,SU95.1}) 

\begin{equation}
T(k) = \frac{\ln [2 e + 1.8 q]}{\ln [2 e + 1.8 q] + q^2 \left[ 14.2 + 731/(1+62.5q)\right]},
\label{EH}
\end{equation} 
where $q \equiv k/\left(\Omega_{\mathrm{m},0} h^2 {\rm Mpc}^{-1}\right)$. The quantity $P(k)$ is the primordial power spectrum of density perturbations, which is linked to the curvature power spectrum (Eq. \ref{eqn:spectrum}) through

\begin{equation}
P(k) = \frac{4}{25} \frac{k^4}{H_0^4\Omega_{\mathrm{m},0}^2}P_\mathcal{R}(k).
\end{equation}
The function $W_R(k)$ is the Fourier transform of the top-hat window function, that can be written as

\begin{equation}
W_R(k) = \frac{3}{(kR)^3}\left[ \sin(kR) - (kR)\cos(kR) \right].
\end{equation}

As can be seen in Figure \ref{fig:rmsMass}, the mass variance for the model with running primordial spectral index is significantly lower than for the concordance model at low masses, while at the massive end the first one is slightly above the second. The transition occurs at $M\sim 3\times 10^{12} M_\odot h^{-1}$, and this kind of trend can be understood as follows. The integral in Eq. (\ref{eqn:variance}) tends to be dominated by large wavenumbers, but it is also suppressed by the Fourier transform of the window function, $W_R(k)$. At low masses the characteristic size $1/R$ of this function is very large, so that the integral is dominated by wavenumbers where the linear power spectrum (see Figure \ref{fig:matter}) in the model with running is lower than for the $\Lambda$CDM model. At high masses instead the window function is narrower, so that the integral is dominated by the part at intermediate wavelengths, where the two linear power spectra are very close to each other, and actually the one in the running model is slightly larger than in the $\Lambda$CDM cosmology. Another consequence of this behavior is that the decrease of $\sigma_M$ as a function of mass is flatter, implying that $\left|d\sigma_M^2/dM\right|$ is also smaller in the model with running. Hence, given Eq. (\ref{eqn:mf}), we expect the mass function to be smaller at low masses in the latter case with respect to the $\Lambda$CDM model, but higher at large masses, where $\sigma_M$ is also larger. In Section \ref{sct:results} we show how this expectation is fulfilled.

\subsection{Halo bias}\label{sct:obs:bias}

The linear bias describes how well a population of cosmic objects traces the underlying, smooth matter density field. It is an ingredient of fundamental importance for computing the expected observed correlation function of galaxies and galaxy clusters, and also for computing the fully non-linear matter power spectrum, as will be more clear further below. In this work, we adopted the bias prescription given by \citet*{SH01.1} as a modification of the original \cite{MO96.1} formula obtained with \citet{PR74.1}-like considerations, that reads 

\begin{eqnarray}\label{eqn:bias}
b(M,z) &=& 1 + a\frac{\Delta_\mathrm{c}}{D_+^2(z)\sigma^2_M} - \frac{1}{\Delta_\mathrm{c}} +
\nonumber\\
&+& \frac{2p}{\Delta_\mathrm{c}} \left[ \frac{[D_+(z)\sigma_M]^{2p}}{[D_+(z)\sigma_M]^{2p} + [\sqrt{a} \Delta_\mathrm{c}]^{2p}} \right].
\end{eqnarray}
Similarly to Eq. (\ref{eqn:mf}), the original \cite{MO96.1} formulation for the bias is obtained by setting $a = 1$ and $p = 0$ in Eq. (\ref{eqn:bias}), while the \citet*{SH01.1} revision is obtained with the values $a = 0.75$ and $p = 0.3$.

The fact that the expression for the linear bias contains the variance of the matter power spectrum implies that some kind of difference is expected when including a running primordial spectral index, as a consequence of the behavior reported in Figure \ref{fig:rmsMass}. Namely, a smaller value of the mass variance is expected to produce a larger bias, and viceversa.

We also note that, both in the mass function and in the halo bias, an impact is also given by the growth factor of density perturbations $D_+(z)$. However, this function does not depend on the initial dark matter density distribution but only on the subsequent expansion history of the Universe.

\subsection{Concentration of cosmic structures}\label{sct:obs:con}

A change in the variance of the matter power spectrum modifies the formation time of cosmic structures \citep*{LI08.1}. The central concentration of dark matter halos reflects the mean matter density at the time of collapse. If structure formation occurs earlier, this matter density would be larger, hence leading to larger average concentrations. It is now well accepted that the average density profiles of dark matter halos follow a universal density profile over many decades in radius and mass, whose shape is effectively independent of cosmology (\citealt*{NA95.1,NA96.1,NA97.1}; \citealt{DO04.1}, see however \citealt{GA08.1}). This density profile (NFW profile henceforth) can be written as the two-parameter function

\begin{equation}
\rho(r) = \frac{\rho_\mathrm{s}}{(r/r_\mathrm{s})(1+r/r_\mathrm{s})^2}.
\end{equation}
The total mass of the halo that is included inside some radius $r$ can be computed as

\begin{equation}\label{eqn:mass}
M(r) = 4\pi \int_0^r \rho(x) x^2 dx = 4\pi \rho_\mathrm{s}r_\mathrm{s}^3G(r/r_\mathrm{s}),
\end{equation}
where the function $G(x)$ takes the well known form

\begin{equation}
G(x) \equiv \ln(1+x) - \frac{x}{1+x}.
\end{equation}

The two parameters $r_\mathrm{s}$ and $\rho_\mathrm{s}$ completely specify the density profile and can be expressed in terms of the virial mass $M$ and concentration $c$. In the remainder of this work we defined the virial mass as the mass contained in the sphere whose mean density equals $\Delta_\mathrm{v} = 200$ times the \emph{average} density of the Universe, that is 

\begin{equation}\label{eqn:massv}
M = \frac{4}{3}\pi R_\mathrm{v}^3 \Delta_\mathrm{v} \rho_\mathrm{m}.
\end{equation}
The virial radius $R_\mathrm{v}$ is the radius of this sphere, and the concentration is defined as the ratio between the virial radius and the scale radius of the profile, $c \equiv R_\mathrm{v}/r_\mathrm{s}$. The scale density can then be related to the concentration of the halo as

\begin{equation}
\rho_\mathrm{s} = \frac{\Delta_\mathrm{v}}{3} \rho_\mathrm{m} \frac{c^3}{G(c)},
\end{equation}
while the dependence of scale radius on the virial mass and the concentration can be written as

\begin{equation}
r_\mathrm{s} = \left( \frac{3M}{4\pi c^3 \Delta_\mathrm{v}\rho_\mathrm{m}} \right)^{1/3}.
\end{equation}

The concentration of a dark matter halo is actually linked to the virial mass according to the hierarchical paradigm for structure formation, since it is expected that small structures collapse earlier and hence are more compact at a given redshift than massive structures. As a consequence the dark matter distribution inside halos depends effectively only on mass and redshift. In order to practically relate the concentration to the virial mass, there exist prescriptions based on the study of samples of high-resolution dark matter halos extracted from numerical simulations of structure formation (\citealt*{NA96.1}; \citealt{BU01.1}; \citealt*{EK01.1}; \citealt{DO04.1,GA08.1}; \citealt*{MA08.3}). All of them share the same generic features, according to which more massive and higher-redshift dark matter halos should be less concentrated than their lower mass, lower redshift counterparts. We, in particular, adopted the prescription given by \citet*{EK01.1}, since it appears to be more general than the others (see \citealt{DO04.1}), and moreover it has been tested for different shapes of the primordial power spectrum, hence it is expected to perform correctly also when a small running is added to the spectral index.

For subsequent use, we also report here the Fourier transform $\hat{\rho}(k,M,z)$ of the halo density profile with respect to radius, that in general can be written as

\begin{equation}\label{eqn:ft}
\hat{\rho}(k,M,z) = 4\pi \int_0^{R_\mathrm{v}} \rho(r,M,z) \frac{\sin(kr)}{kr}r^2dr.
\end{equation}
This definition conveniently implies $\hat{\rho}(0,M,z) = M$. The Fourier transform of the NFW density profile can be expressed analitically, according to 

\begin{eqnarray}
\hat{\rho}(k,M,z) &=& 4\pi \rho_\mathrm{s}r_\mathrm{s}^3 \left\{\frac{}{}\sin(kr_\mathrm{s}) \left[\mathrm{Si}((1+c)kr_\mathrm{s}) - \mathrm{Si}(kr_\mathrm{s}) \right] \right. + 
\nonumber\\
&+&\cos(kr_\mathrm{s}) \left[\mathrm{Ci}((1+c)kr_\mathrm{s}) - \mathrm{Ci}(kr_\mathrm{s}) \right] -
\nonumber\\
&-&\left.\frac{\sin(kr_\mathrm{s}c)}{(1+c)kr_\mathrm{s}} \right\}
\end{eqnarray}
(\citealt*{RU08.2}; \citealt{SC01.1}), where $\mathrm{Si}(x)$ and $\mathrm{Ci}(x)$ are the sine and cosine integrals respectively. Please note that, for $x \rightarrow 0$, $\mathrm{Si}(x) \simeq 0$ while $\mathrm{Ci}(x) \simeq \ln(x)$, thus we have that $\hat{\rho}(0,M,z) = 4\pi \rho_\mathrm{s}r_\mathrm{s}^3 G(c)$, that according to Eq. (\ref{eqn:mass}) correctly equals the virial mass of the halo.

\subsection{Matter power spectrum and Baryon Acoustic Oscillation}\label{sct:obs:bao}

The Baryon Acoustic Oscillation (BAO henceforth) is a wiggle that show up in the matter power spectrum at a comoving scale $k \sim 0.1 h$ Mpc$^{-1}$. It is a relic of the acoustic peaks that are seen in the angular spectrum of temperature fluctuations in the CMB, due to oscillations of the baryon-photon fluid within dark matter potential wells prior to recombination. The BAO signature has already been detected in large galaxy surveys such as the SDSS and 2dFGRS \citep{EI05.1,PE07.2,PE07.1}, and constitutes one of the most promising cosmological probes for the future.

The BAO enter the matter power spectrum in the linear part, but is subsequently modified by the non-linear evolution, that couples modes at different wavenumbers \citep{CR06.2,CR06.1,MA07.3,MA08.2}. Therefore we computed the fully non-linear power spectrum, which has also been necessary for estimating the weak lensing power spectrum described in the next Section \ref{sct:obs:wl}, using the semi-analytic halo model \citep{MA00.3,SE00.1}. In the halo model, the power spectrum is set by the sum of two terms,

\begin{equation}
P_\mathrm{NL}(k,z) = P_1(k,z)+P_2(k,z).
\end{equation}
The first term dominates on large scales and is given by dark matter particles residing in different halos, hence it depends on the correlations between individual halos. The second term dominates on the smallest scales and takes into account particles that are included in the same halo, hence it is extremely sensitive to the inner structure of halos themselves (Section \ref{sct:obs:con}). Here we report only the main features of the two terms composing the matter power spectrum, deferring to \citet*{CO00.2,SE00.1,RE02.2,SE03.2,BE10.1,FE10.1} for a full discussion.

Let $P(k,z)=T^2(k)P(k)D_+^2(z)$ be the linear power spectrum of density fluctuations extrapolated at redshift $z$. The two terms making up the fully non-linear power spectrum can be written as

\begin{equation}\label{eqn:p1}
P_1(k,z) = \int_0^{+\infty} n(M,z) \left[\frac{\hat{\rho}(M,z,k)}{\rho_\mathrm{m}}\right]^2 dM
\end{equation}
and
\begin{equation}\label{eqn:p2}
P_2(k,z) = \left[ \int_0^{+\infty} n(M,z)b(M,z)\frac{\hat{\rho}(M,z,k)}{\rho_\mathrm{m}}dM \right]^2 P(k,z).
\end{equation}
Two non-trivial constraints have to be taken into account when performing the computation of the two terms above, both of which are to be enforced numerically. They read

\begin{equation}\label{eqn:c1}
\int_0^{+\infty} n(M,z) \frac{M}{\rho_\mathrm{m}} dM = 1
\end{equation}
and

\begin{equation}\label{eqn:c2}
\int_0^{+\infty} n(M,z)b(M,z)\frac{M}{\rho_\mathrm{m}}dM = 1.
\end{equation}
The first constraint simply imposes that all the matter in the Universe is included within halos of some mass. As can be easily verified, this constraint is fulfilled by the \cite{PR74.1} mass function, and also the \cite{SH02.1} mass function is normalized such to satisfy Eq. (\ref{eqn:c1}). The second constraint requires that on very large scales the non-linear matter power spectrum approaches the linear one.

Particular care must be taken about the average relation between the concentration and the virial mass of dark matter halos, that has an important impact on the very non-linear part of the power spectrum. In particular, choosing the average concentration-mass relation measured in samples of high-resolution simulated halos, such as the prescription of \citet*{EK01.1}, does not return a power spectrum in agreement with the one measured in large scale cosmological simulations \citep{PE94.1,SM03.1}, more so for higher redshifts. This can be due either to assuming average density profiles for dark-matter halos and average relations between structural properties thereof, that do not fully capture the true complexity of cosmic structures, or to extrapolating fits to numerical simulations at unprobed very small scales. As a matter of fact, many authors choose their own concentration-mass relation in order to reproduce the fully non-linear matter power spectrum of numerical simulations (\citealt{CO00.2,RE02.2}; \citealt*{BE10.1}), irrespective of the fact that such relation is also observed in samples of high-resolution halos. Either way, we have chosen to follow \citet{FE10.1}, by letting the concentration drop more steeply with redshift than predicted by \citet*{EK01.1}. In this way we obtain fair agreement with the results of the public \texttt{halofit} code \citep{SM03.1}.

\subsection{Weak lensing power spectrum}\label{sct:obs:wl}

Apart from tracers of the large scale structures such as galaxies, information on the matter power spectrum can be obtained through cosmological weak gravitational lensing. The power spectrum of the shear signal that can be derived by averaging the distortion of images of high-redshift galaxies is an integral of the fully non-linear matter power spectrum, weighted with the source redshift distribution.

The power spectrum of the lensing convergence can be written as \citep{BA01.1}

\begin{equation}\label{eqn:wl}
C_\ell = \frac{9}{4}H_0^2\Omega_{\mathrm{m},0}^2 \int_0^{\chi_\mathrm{H}} P\left(\frac{\ell}{f_K(\chi)},\chi \right) \frac{W^2(\chi)}{a^2(\chi)} d\chi,
\end{equation}
where $\chi = \chi(z)$ is the comoving distance out to redshift $z$, $a$ is the scale factor normalized to unity today and $f_K(\chi)$ is the comoving angular-diameter distance corresponding to the comoving distance $\chi$, which depends on the spatial curvature $K$ of the Universe. The integral in Eq. (\ref{eqn:wl}) extends formally out to the horizon size $\chi_\mathrm{H}$, however the integrand becomes zero well before this limit is reached, due to the absence of sources at $z \gtrsim 4$. The redshift distribution of sources $n(z)$ has a fundamental role in the evaluation of the weak lensing power spectrum, as it defines the integration kernel according to

\begin{equation}
W(\chi) = \int_\chi^{\chi_\mathrm{H}} n(\chi') \frac{f_K(\chi-\chi')}{f_K(\chi')} d\chi'.
\end{equation}
The Eq. (\ref{eqn:wl}) for the convergence power spectrum was obtained using Fourier expansion under the Limber's approximation \citep{BA01.1}, while the exact expression would make use of spherical harmonics. However, it was recently shown by \cite{JE09.1} that, at least when considering only the convergence power spectrum, the accuracy of the Limber's approximation is very good, better than $1\%$ at $\ell>10$, corresponding to $2\pi/\ell \lesssim 2 \times 10^3$ arcmin.

\begin{figure}
	\includegraphics[width=\hsize]{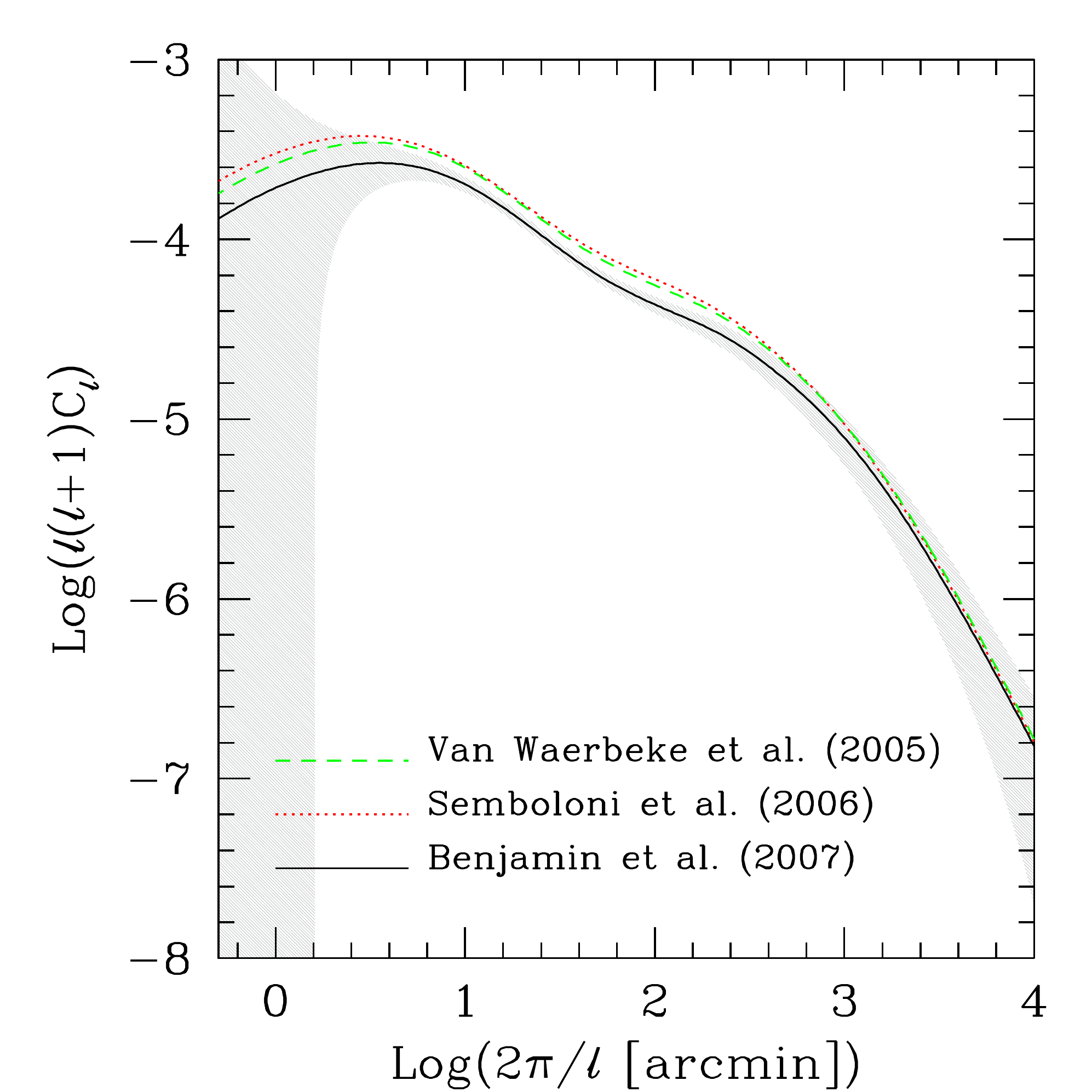}\hfill
	\caption{The weak lensing power spectra as computed in the $\Lambda$CDM model for three different redshift distribution of sources, as labeled in the plot. The shaded area represents the statistical error for the power spectrum computed with the \citet{BE07.2} redshift distribution, and assuming a EUCLID-like weak lensing survey.}
	\label{fig:weakLensing}
\end{figure}

Several choices for the redshift distribution of background sources to be adopted for cosmic shear studies are available in the literature. One of the most recent ones is presented in the work of \cite{BE07.2}, where a detailed analysis of the photometric redshift distribution in four different fields is reported. The four fields considered were the CFHTLS wide survey \citep{VA02.1,HO06.1}, the GaBoDS \citep{HE07.1} field, the VIRMOS-DESCART \citep{VA01.1,MC03.1,LE04.1} project, and the RCS \citep{HO02.1} survey. \cite{BE07.2} fit the photometric redshift distribution in the four fields using the three-parameter formula

\begin{equation}\label{eqn:z1}
n(z) = \frac{\beta}{z_0\Gamma\left[(1+\alpha)/\beta\right]} \left( \frac{z}{z_0} \right)^\alpha \exp \left[ -\left( \frac{z}{z_0} \right)^\beta \right].
\end{equation}
The same fitting formula has also been used by \cite{VA05.2} and \cite{SE06.1} to fit the photometric redshift distribution of the Hubble Deep Field (HDF). As noted by \cite{BE07.2} and suggested by \cite{VA06.1} however, the HDF suffers of sample variance, and maybe also subject to a selection bias.
\cite{BE07.2} noted that the formula in Eq. (\ref{eqn:z1}) does not fit very well their photometric redshift distributions if all galaxies are included, and proposed a different functional for that performed a better fit. However, when considering only their high-confidence redshift interval (outside which the fraction of catastrophic errors reaches $40-70\%$), Eq. (\ref{eqn:z1}) becomes a good fit. We chose to stick to this choice, and for each of the three parameters in Eq. (\ref{eqn:z1}) we adopted the mean of the values for the four fields analyzed by \cite{BE07.2}.

\begin{figure}
	\includegraphics[width=\hsize]{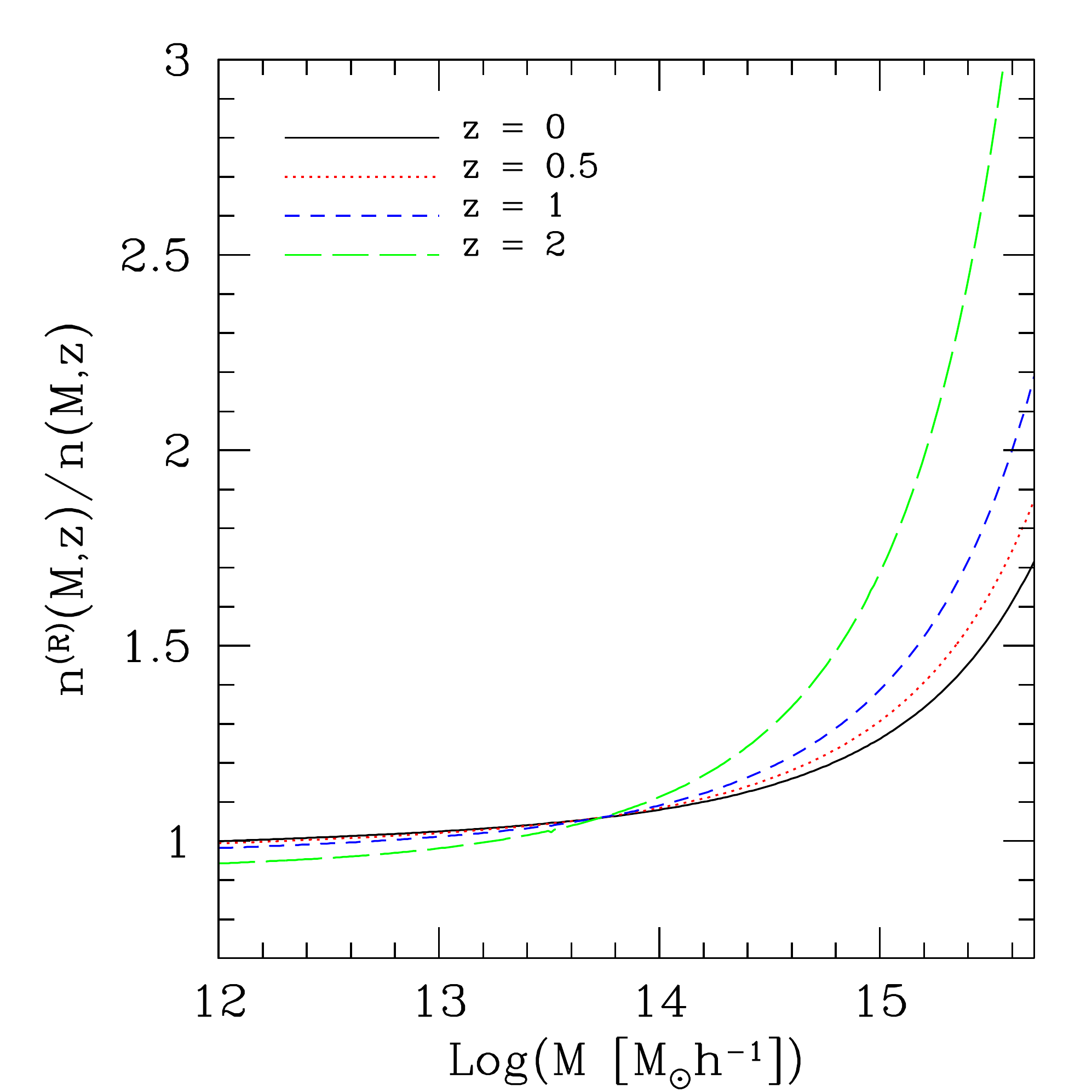}\hfill
	\caption{The ratio of halo mass function in the model with running of the primordial spectral index  to the mass function in the $\Lambda$CDM cosmology. The structure abundance is computed according to the \citet{SH02.1} prescription, and four different redshift are shown, as labeled in the plot.}
	\label{fig:massFunctionRatio}
\end{figure}

In order to show the effect of different choices for the background source redshift distribution, in Figure \ref{fig:weakLensing} we report the weak lensing power spectrum as obtained in the $\Lambda$CDM model for the distributions of  \cite{VA05.2}, \cite{SE06.1}, and \cite{BE07.2} (our fiducial one). As can be seen, the difference between different source redshift distributions can be quite significant, especially at intermediate scales, implying that the choice of the redshift distribution must be carefully addressed, given the precision level reached by future surveys.

The statistical error presented in Figure \ref{fig:weakLensing} has been evaluated according to \cite{KA92.1,KA98.1,SE98.1,HU02.1}, using the prescription

\begin{equation}\label{eqn:er}
\Delta C_\ell = \sqrt{\frac{2}{(2\ell+1)\Delta\ell f_\mathrm{sky}}} \left( C_\ell + \frac{\gamma^2}{\bar{n}} \right).
\end{equation}
In Eq. (\ref{eqn:er}) $\bar{n}$ is the average number density of galaxies in the survey at hand, $f_\mathrm{sky}$ is the survey area in units of the sky area, and $\gamma$ is \emph{rms} intrinsic shape noise for each galaxy. For an EUCLID-like future weak lensing survey \citep{LA09.1} we adopted $\bar{n} = 40$ arcmin$^{-2}$, $f_\mathrm{sky} = 0.5$ and $\gamma = 0.22$ \citep*{ZH09.1}. The parameter $\Delta\ell$ represents the bin width over which the spectrum is practically averaged. We adopted $\Delta\ell=1$ henceforth \citep{TA07.1,TA09.1}.

\subsection{Integrated Sachs-Wolfe effect}\label{sct:obs:isw}

CMB photons traveling toward Earth pass through the LSS that is evolving due to gravitational instability. At low redshift, when the expansion of the Universe becomes dominated by the cosmological constant and the density fluctuations evolve non-linearly, potential wells associated with overdense regions can evolve significantly during the time a photon travels through them. The photon acquires a gravitational blueshift when entering the potential well, and a redshift when exiting it. If the potential well has deepened in the meanwhile, the magnitude of the latter is larger than that of the former, hence the photon undergoes a net energy loss. The opposite process obviously happens when the photon crosses potential hills, due to underdense regions in the cosmic web.

This is the Integrated Sachs-Wolfe (ISW henceforth, see \citealt{SA67.1}) effect, whose observable signatures are the occurrence of secondary temperature anisotropies of the CMB that are correlated with the LSS in the Universe. This causes an additional contribution to the CMB spectrum, that however manifests only at very large scales (low multipoles $\ell$), since the effect is driven by the nonlinear evolution of potential fluctuations, that occurs at very low redshift in a dark energy-dominated cosmological model. It is expected that a scale dependence in the primordial spectral index would modify the LSS at low redshift, and hence would produce a different ISW signal.

\section{Results}\label{sct:results}

In this section we report the results concerning all the different observables that have been introduced in Section \ref{sct:observables}, for the two fiducial cosmological models detailed in Section \ref{sct:cosmoparams}.

\subsection{Galaxy cluster abundance}\label{sct:res:abundance}

\begin{figure*}
	\includegraphics[width=0.45\hsize]{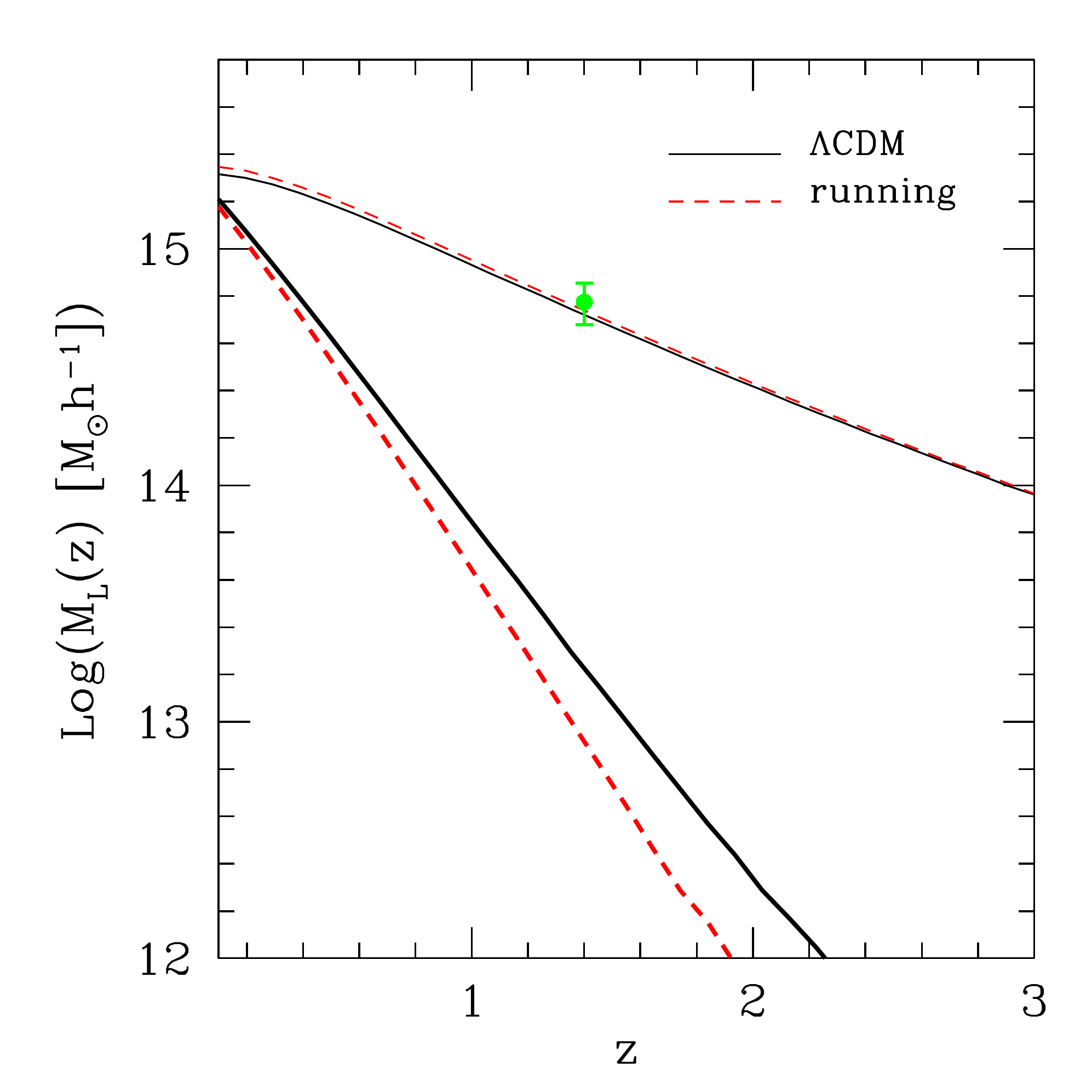}\hfill
	\includegraphics[width=0.45\hsize]{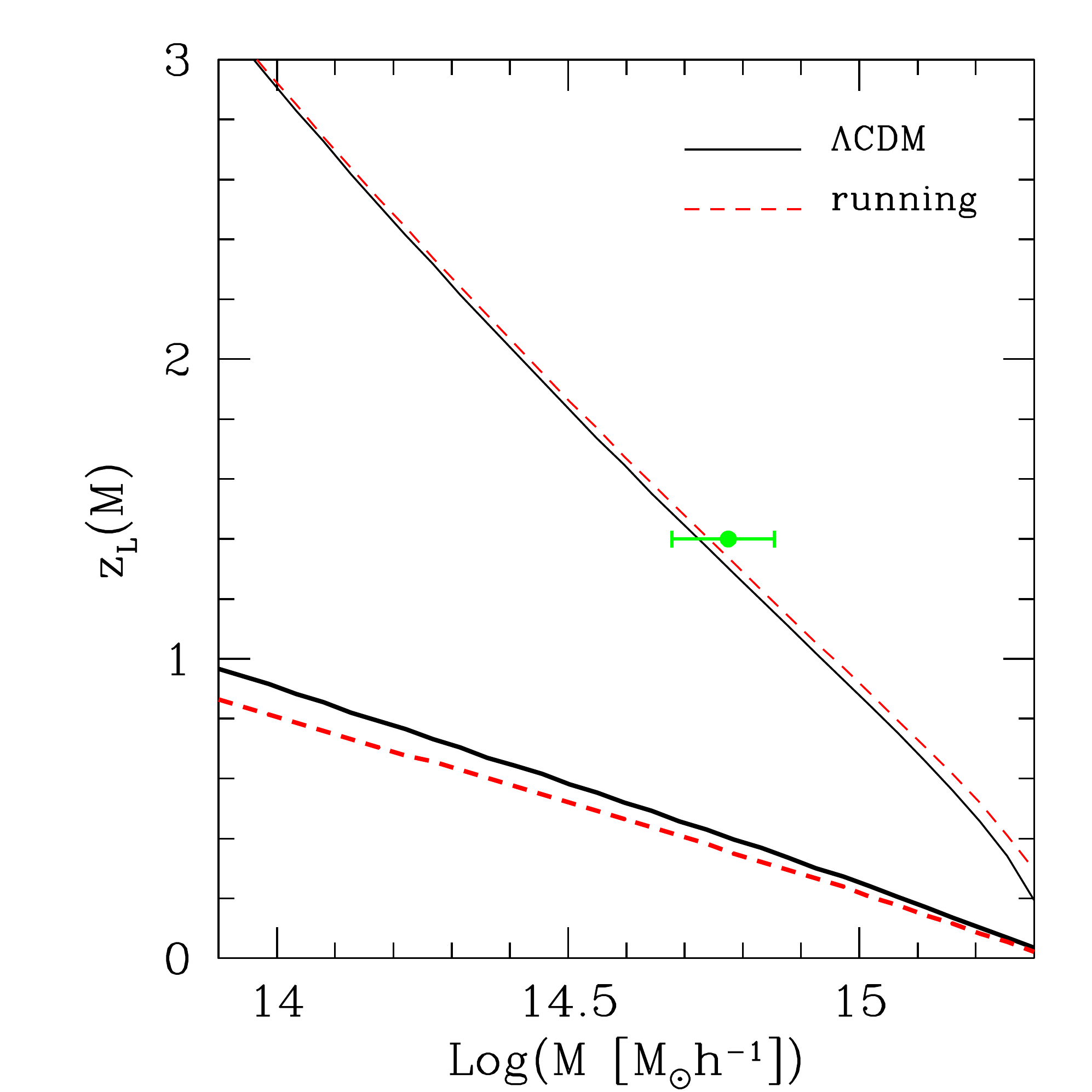}
	\caption{\emph{Left panel}. The minimum mass $M_\mathrm{L}(z)$ such that the number of objects in the whole sky with redshift larger than $z$ and mass larger than $M_\mathrm{L}(z)$ is $N_*=10$ (thin lines). Results are shown for both models  considered in this work, as labeled in the plot, and for comparison the thick lines show $10^3M_*(z)$, where $M_*(z)$ is the characteristic non-linear mass at redshift $z$. \emph{Right panel}. The limiting redshift $z_\mathrm{L}(M)$ such that the number of objects in the whole sky with mass larger than $M$ and redshift larger than $z_\mathrm{L}(M)$ equals $N_*$ (thin lines). The thick lines represent the redshift at which the non-linear mass $10^3M_*(z)$ equals the value reported on the abscissa. Black solid and red dashed lines are as in the left panel. In both panels, the green points with errorbars represent the mass range allowed for the massive cluster XMMUJ$2235.3-2557$, at redshift $z = 1.4$.}
	\label{fig:rareEvents}
\end{figure*}

In Figure \ref{fig:massFunctionRatio} we report the ratio of the mass function in the model with running of the primordial spectral index to the mass function in the $\Lambda$CDM cosmology for four different values of the redshift. As can be seen, the abundance of objects in the former model is smaller than in the latter at low masses, while it can be significantly larger at high masses. These deviations are amplified with increasing redshift, as one might naively expect. This kind of behavior is in agreement with the discussion about the trend of $\sigma_M$ in Section \ref{sct:obs:abundance}, and with the expectations presented there. The increment in the mass function due to the running can be as large as a factor of $2-3$ for very high masses and large redshifts. However, it should be kept in mind that this increment is not solely due to the running itself, but also to the fact that, e.g., the normalization $\sigma_8$ in the model with running is slightly larger than in the $\Lambda$CDM cosmology.

Since the mass function represents only a space density of objects, the plot in Figure \ref{fig:massFunctionRatio} is not very informative, in the sense that no structures are expected to be found in the sky with very high mass and high redshift, where the effect of running should be larger. Motivated by this fact, and by the recent discussion that arose in the literature regarding the occurrence of rare events and the indications that these can give on the level of non-Gaussianity of the primordial density field \citep{JI09.1,SA10.1}, we briefly addressed the issue of what a "rare event" meaningfully is and how the abundance thereof is affected by a running of the spectral index. Generically, a rare event is the occurrence of a high-mass galaxy cluster at high redshift, but what "high" exactly means remains subjective.

In Figure \ref{fig:rareEvents}, we try to specify this concept a little bit more clearly. In the left panel we report, for both the fiducial models adopted in this work, the limiting mass $M_\mathrm{L}(z)$ defined such that the number of structures in the entire sky with redshift larger than $z$ and mass larger than $M_\mathrm{L}(z)$ equals $N_*=10$. The choice of the number $N_*$ is arbitrary to some extent, however we believe it fairly characterizes rare objects. As comparison we also report in the same panel the redshift evolution of the non-linear mass $10^3 M_*(z)$, defined such that $D_+(z)\sigma_{M_*}=\Delta_\mathrm{c}$.
We first of all note that both these mass scales decrease with redshift, in both cosmological models. This is to be expected because massive objects are more rare at higher redshift, and the growth factor $D_+(z)$ is a decreasing function of redshift. Secondly, the limiting mass $M_\mathrm{L}(z)$ for the model with running of the primordial spectral index is always slightly higher than the corresponding mass for the concordance model because the mass function in the former cosmology is always higher than for the latter model, hence it is easier to reach $N_*$ objects in the whole sky. The non-linear mass $M_*(z)$ drops much more steeply with redshift than $M_\mathrm{L}(z)$, and its value in the $\Lambda$CDM cosmology is always larger than in the cosmology with running spectral index, except at $z\sim 0$. This can be understood since at $z > 0$ we have $D_+(z)< 1$, and according to Figure \ref{fig:rmsMass} a fixed $\sigma_M>\Delta_\mathrm{c}$ is reached for lower masses in the running model as compared to the $\Lambda$CDM cosmology. The two trends of $\sigma_M$ with mass also easily explain why the differences in $M_*(z)$ between the two cosmologies grow with redshift.

The right panel of Figure \ref{fig:rareEvents} is similar to the left one, but with mass and redshift having exchanged roles. Namely we plot, for both the fiducial cosmologies considered here, the limiting redshift $z_\mathrm{L}(M)$ such that the number of objects in the entire sky with mass larger than $M$ and redshift larger than $z_\mathrm{L}(M)$ equals $N_*$. Additionally, we also plot the redshift at which the non-linear mass scale $10^3M_*(z)$ equal a fixed value $M$ of the mass. In this case as well, the redshift at which it is necessary to push in order to have, for a given mass, less than $N_*$ objects in the whole sky is higher for the model with running of the primordial spectral index than for the $\Lambda$CDM model, as expected. 

In the left and right panels of Figure \ref{fig:rareEvents} we also report the redshift and mass range estimated for the high-$z$ cluster XMMUJ$2235.3-2557$, whose high mass was recently estimated through weak gravitational lensing by \cite{JE09.2} (see also \citealt{RO09.1}). As can be seen, this cluster falls almost perfectly on the curves for $M_\mathrm{L}(z)$ and $z_\mathrm{L}(M)$, hence we expect to find $\sim 10$ such clusters in the whole sky, and it qualifies as a "rare event" according to our definition. This conclusion is in agreement with \cite{JI09.1}, where the authors state a number of $7$ in the whole sky for this kind of objects, by using a cosmology with a lower value of $\sigma_8$ than our own.

\begin{figure*}
	\includegraphics[width=0.45\hsize]{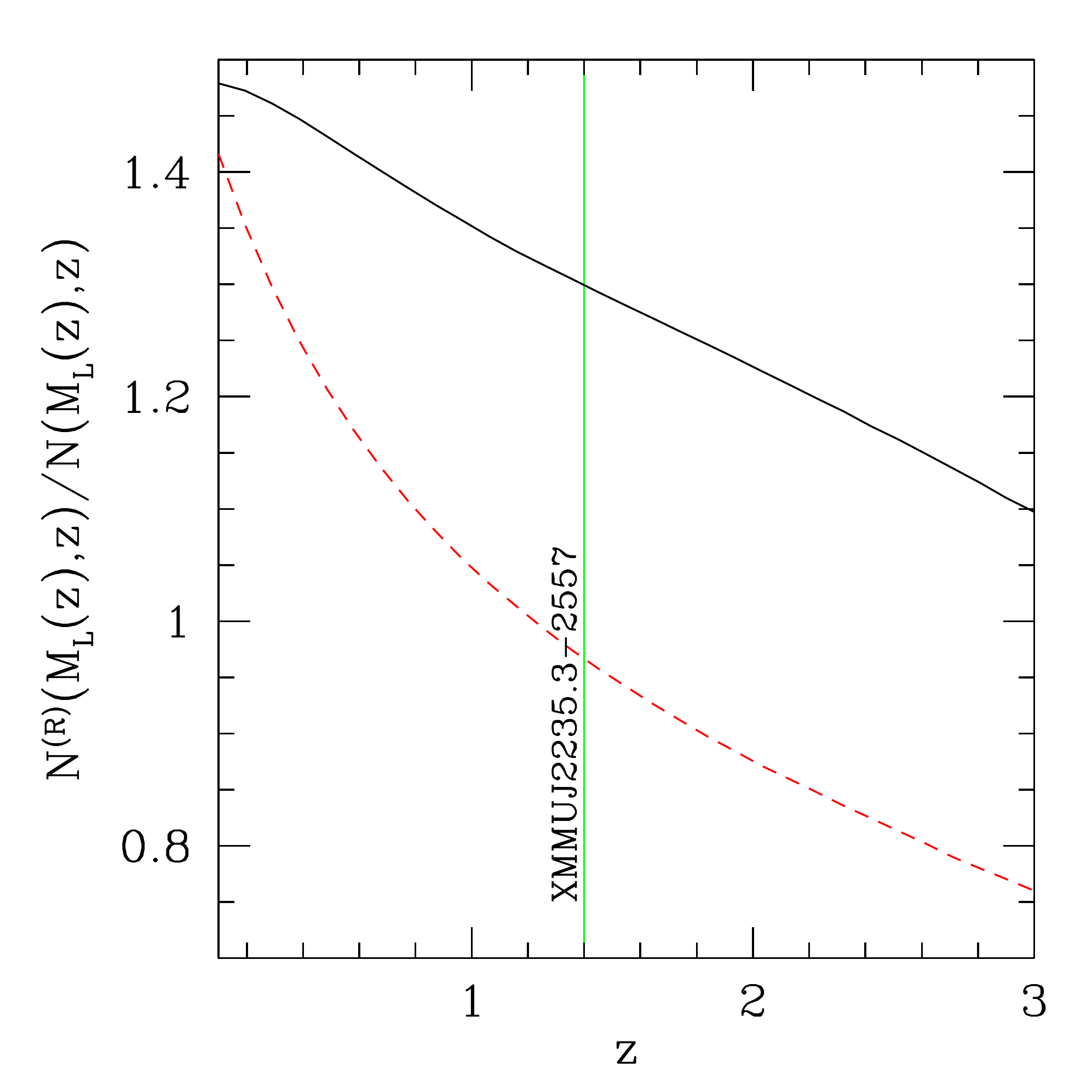}\hfill
	\includegraphics[width=0.45\hsize]{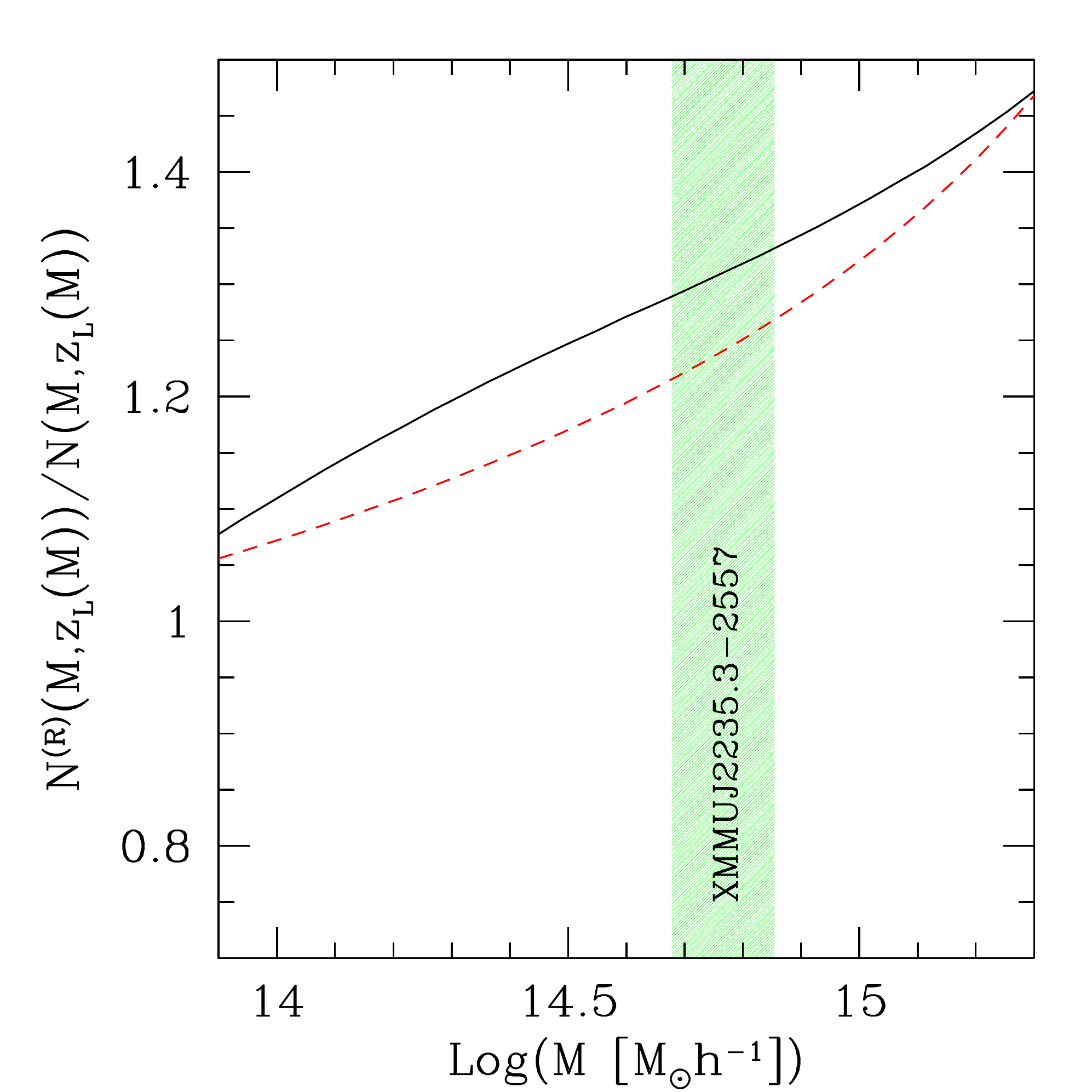}
	\caption{\emph{Left panel}. The ratio between the number of objects, computed in the model with running of the spectral index, with redshift larger than $z$ and mass larger than the limiting mass $M_\mathrm{L}(z)$ to the same quantity computed in the concordance model (black solid line), where $M_\mathrm{L}(z)$ was always referred to the standard model. The red dashed line represents the same quantity computed by replacing $M_\mathrm{L}(z)$ with $10^3M_*(z)$ for the $\Lambda$CDM cosmology. The vertical green solid line indicates the redshift of the massive cluster XMMUJ$2235.3-2557$. \emph{Right panel}. The ratio between the number of objects, computed in the model with running of the spectral index, with mass larger than $M$ and redshift larger than the limiting redshift $z_\mathrm{L}(M)$ to the same quantity computed in the concordance model (black solid line), where $z_\mathrm{L}(M)$ was always referred to the standard model. The red dashed line represents the same quantity computed by replacing $z_\mathrm{L}(M)$ with the redshift at which $10^3M_*(z) = M$ for the $\Lambda$CDM cosmology. The vertical green shaded region indicates the allowed mass range for the cluster XMMUJ$2235.3-2557$.}
	\label{fig:rareAbundance}
\end{figure*}

In the left panel of Figure \ref{fig:rareAbundance} we show the ratio of the number of objects, computed in the model with running of the spectral index, that have mass larger than $M_\mathrm{L}(z)$ and redshift larger than $z$ to the same quantity evaluated in the $\Lambda$CDM cosmology, as a function of redshift. Here $M_\mathrm{L}(z)$ is always estimated in the concordance model. As a reference, we also show the same quantity computed when replacing the limiting mass with $10^3M_*(z)$. In the right panel of the same Figure we report the same ratio but with mass and redshift having exchanged roles, as in the right panel of Figure \ref{fig:rareEvents}. It is apparent that the abundance of objects similar to the cluster XMMUJ$2235.3-2557$ is increased by $\sim 30\%$ in the model with running of the spectral index. The abundance of objects as rare as that but with lower redshift can be augmented by as much as $\sim 45\%$. We also note that, since the non-linear mass $10^3M_*(z)$ is much smaller than the limiting mass $M_\mathrm{L}(z)$, if the former is adopted in order to define rare objects then the increment due to the running of the spectral index is less pronounced, as one might expect. Generically, the fact that the effect of running on rare objects is more evident at the highest mass is quite natural. That the same effect is larger at lower redshift is maybe less intuitive, since the mass function ratio reported in Figure \ref{fig:massFunctionRatio} increases with redshift. However, this ratio also increases with increasing mass, and the limiting masses adopted also becomes larger at lower redshift (see Figure \ref{fig:rareEvents}). Thus we can conclude that in this case the increase in the mass function due to the larger mass overcomes the decrease thereof due to the lower redshift. 

\begin{figure}
	\includegraphics[width=\hsize]{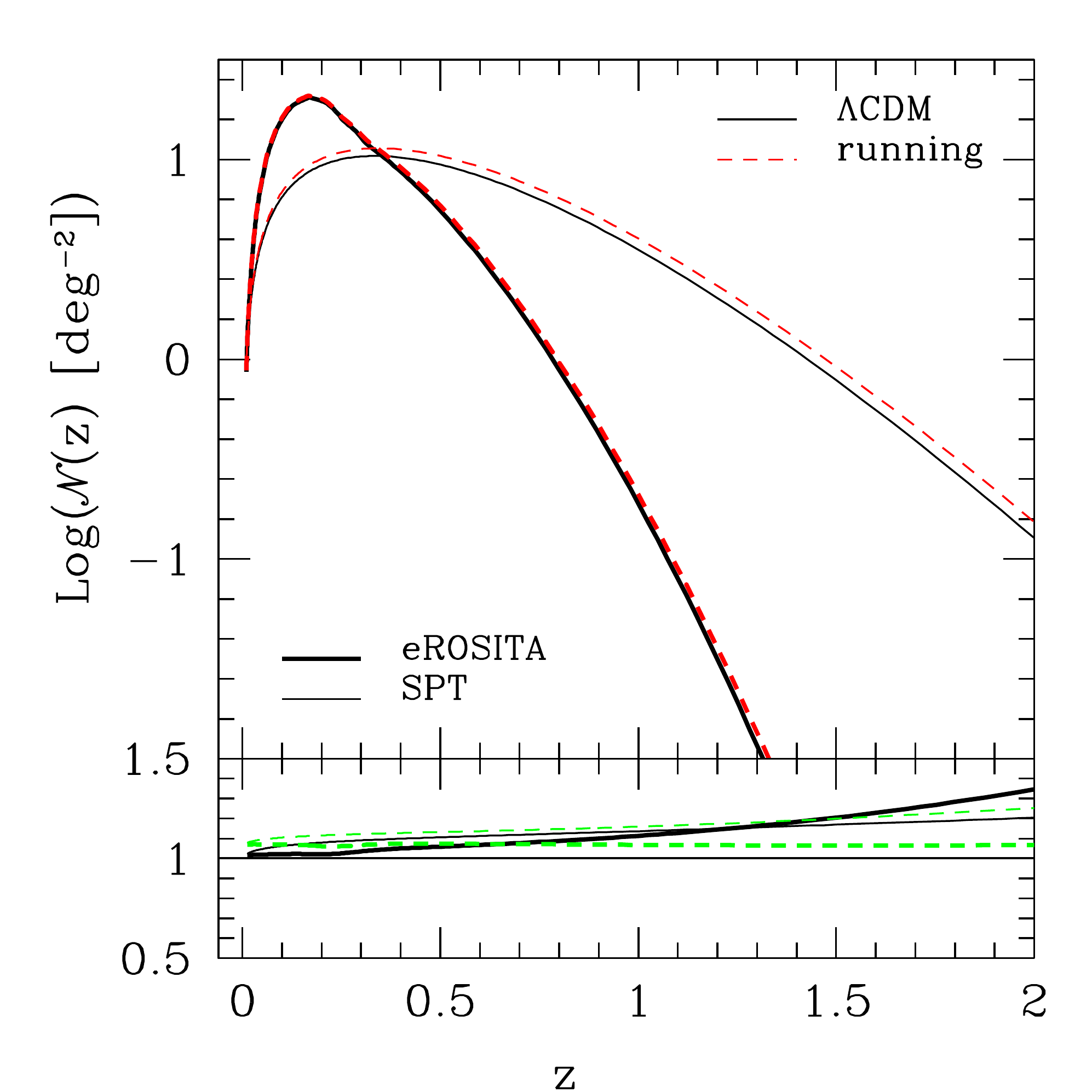}\hfill
	\caption{The redshift distributions predicted for the cluster catalogues obtained by \emph{e}ROSITA (thick lines) and SPT (thin lines) for the two cosmological models under consideration, as labeled in the plot. In the lower inset black solid lines show the ratio of the two distributions to the respective $\Lambda$CDM case, while the green dashed lines refer to the same quantity evaluated by setting $\alpha_{\mathrm{S},0}=0$ in the running model (see the text for details).}
	\label{fig:clusterAbundance}
\end{figure}

The increment in the abundance of rare objects due to the running spectral index that is displayed in Figure \ref{fig:rareAbundance} is overall quite modest. A $\sim 30\%$ increase in number counts is at the level of the Poisson noise for uncommon events. Besides, uncertainties in the mass determination for high-redshift clusters contributes to render this method likely not useful for the purpose of the present work.

As a final step concerning cluster abundance, in Figure \ref{fig:clusterAbundance} we show the predictions for the redshift distributions of the galaxy cluster catalogues that will be produced by the X-ray satellite \emph{e}ROSITA\footnote{\sc http://www.mpe.mpg.de/projects.html\#erosita} and the submm facility South Pole Telescope\footnote{\sc http://pole.uchicago.edu/} (SPT henceforth). The two redshift distributions were obtained in a way similar to the one already described in \citet*{FE09.2,FE09.1}, which consists in combining simple scaling relations between cluster mass and observables with the sensitivity limits expected for the two surveys. For \emph{e}ROSITA we adopted a limiting X-ray flux of $F_\mathrm{lim} = 3.3 \times 10^{-14}$ erg s$^{-1}$ cm$^{-2}$ in the energy band $[0.5,2]$ keV, as specified in the mission definition document, and a relation between mass and bolometric luminosity that reads

\begin{equation}
L(M,z) = 3.087 \times 10^{44} \mathrm{erg}\,\mathrm{s}^{-1} h^{-2} \left[ \frac{M}{10^{15}M_\odot}h(z) \right]^{1.554}.
\end{equation}
The former scaling relation is a combination of the virial relation between mass and X-ray temperature, with normalization based on the simulations of \cite{MA01.1}, with the luminosity-temperature relation found by \cite{AL98.1}. This relation turns out to reproduce well the scaling relation measured for the REFLEX cluster sample by \cite{RE02.1}. The bolometric luminosity was then converted into a band flux by modeling the intracluster plasma via a \cite{RA77.1} plasma model with metallicity $Z=0.3Z_\odot$ \citep{FU98.1,SC99.1} and implemented with the \texttt{xspec} software package \citep{AR96.1}. For SPT, we adopted a limiting SZ flux density at $\nu_0 = 150$ GHz of $S_{\nu_0,\mathrm{lim}} = 5$ mJy \citep{MA03.1} and a simulation-calibrated scaling relation given by \citet{SE07.1}, that reads

\begin{equation}
S_{\nu_0}(M,z) = \frac{2.592\times 10^8 \mathrm{mJy}}{(D_\mathrm{A}(z)/1 \mathrm{Mpc})^2} \left( \frac{M}{10^{15}M_\odot} \right)^{1.876} E(z)^{2/3},
\end{equation}
where $D_\mathrm{A}(z)$ is the angular diameter distance out to redshift $z$ and $E(z) \equiv h(z)/h$.

The difference between the two catalogues is evident, due to the different selection function that X-ray and SZ observations imply (see the discussion in \citealt*{FE09.2}). It can also be seen that in the model with running of the spectral index the number of objects is always higher than the one for the concordance cosmology. This is an obvious consequence of the differences in the mass function at cluster scales that were discussed above. The differences between the two cosmologies are larger for the \emph{e}ROSITA catalogue than for the SPT one at $z \gtrsim 1$. This is so because at high-redshift the limiting mass of the latter is significantly smaller than that of the former. However, at those redshifts very few objects are present in the \emph{e}ROSITA catalogue, so that the latter is going to be largely dominated by low-redshift objects. Since this is true only to a lesser extent in the SPT catalogue, the differences in the total cluster number counts between the two cosmologies are expected to be more pronunced in the latter than in the former. The green dashed lines in the lower inset of Figure \ref{fig:clusterAbundance} show what happens if we allow the variations on the cosmological parameters induced by the presence of spectral running, but set the running itself to zero. In other words, we isolated the effect of the cosmological parameters different from $\alpha_{\mathrm{S},0}$. In this case the modifications to the mass function depicted in Figure \ref{fig:massFunctionRatio} are substantially different, bringing a larger increase at low masses and a reduction of the effect at high masses. As a consequence the redshift distribution for SPT, that is more sensitive at low masses, is slightly increased, while for \emph{e}ROSITA, that samples very high masses especially at high redshift, the effect is reduced. This result shows that the effect of spectral running on cluster counting is substantial even if all the other cosmological parameters are held fixed. We would like however to emphasize that the shift in the other cosmological parameters is a consequence of the assumption on existence of running, hence only the total, combined effect should be considered in this analysis.

Integrating over the redshift distributions presented in Figure \ref{fig:clusterAbundance} and assuming realistic values of $f_\mathrm{sky} = 0.5$ for \emph{e}ROSITA and $f_\mathrm{sky} = 0.2$ for SPT we found an increment  in the total number of objects in the catalogues of only $\sim 4\%$ in the first case and $\sim 11\%$ in the second. Although both these increments are going to be well above the Poisson noise, given the very large number of clusters that are to be detected by these surveys, SPT is clearly more appealing for finding signatures of running through cluster counting. The situation of X-ray-based surveys might improve for future missions that are currently being planned, such as the Wide Field X-ray Telescope (WFXT, see \citealt*{MU10.1}). If we assume to have also a broad redshift information, then we may think of splitting the two catalogues in a low and a high redshift part. The redshifts cutting the higher-$z$ half of the two samples would be $z_\mathrm{h} \simeq 0.24$ for \emph{e}ROSITA and $z_\mathrm{h} \simeq 0.51$ for SPT. Then, in the high redshift sample we would find an increment due to the running of $\sim 5 \%$ in the first case and $\sim13\%$ in the second one.

\subsection{Halo bias}\label{sct:res:bias}

\begin{figure}
	\includegraphics[width=\hsize]{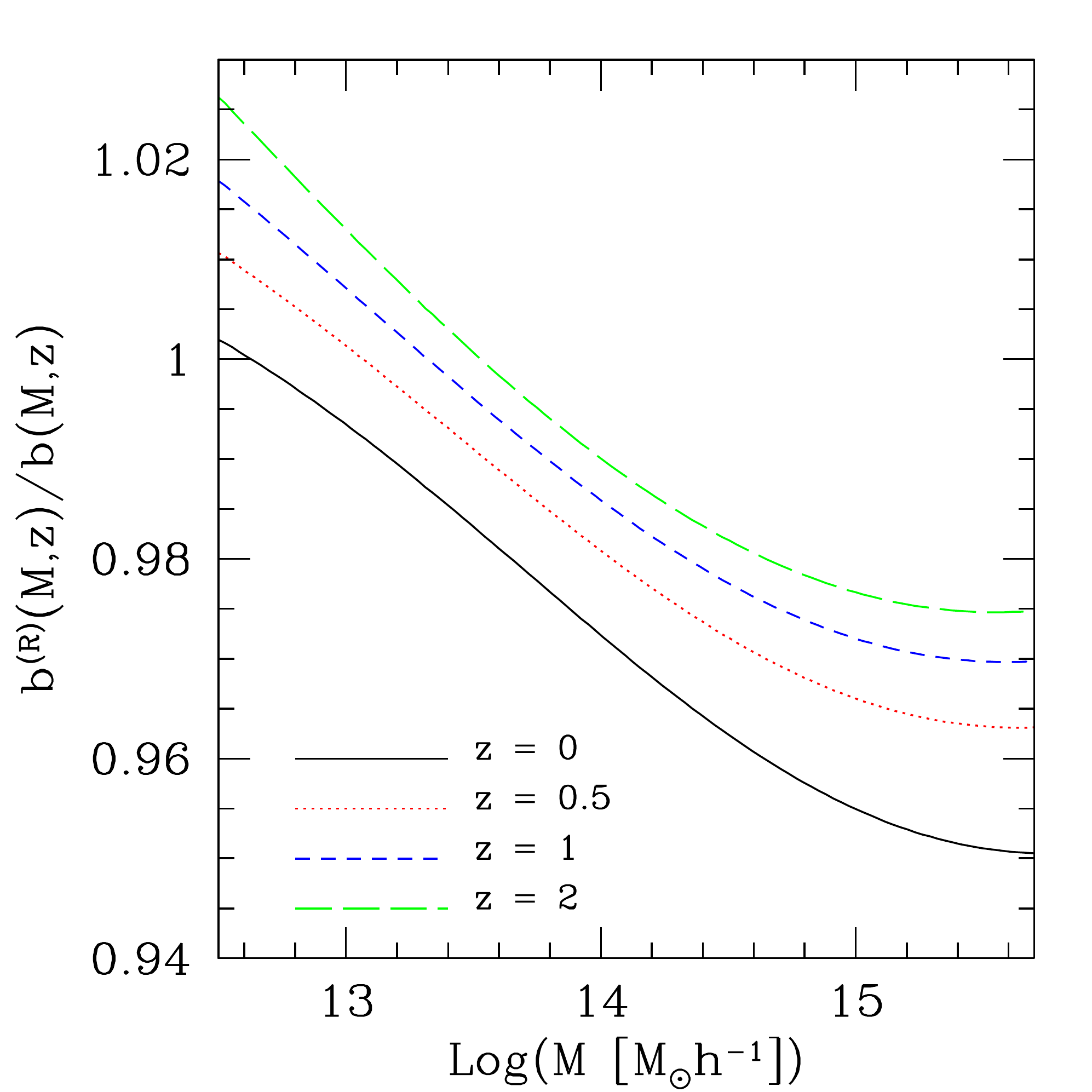}\hfill
	\caption{The ratio between the halo bias in the model with running primordial spectral index to the same quantity in the reference $\Lambda$CDM cosmology, as a function of mass and for the different redshifts labeled in the plot.}
	\label{fig:haloBiasRatio}
\end{figure}

Since the linear bias of dark matter halos also depends on the mass variance, we expect that it is modified when the power spectrum of primordial density fluctuations is not a perfect power-law. In Figure \ref{fig:haloBiasRatio} we report the ratio between the halo bias in the model with running of the spectral index and the same quantity in the reference $\Lambda$CDM cosmology. Results are shown as a function of mass and for several different redshifts. As can be seen, the halo bias in the former model is generically smaller than in the reference model at the high mass end. This is to be expected since in this regime the structure abundance is also augmented by the spectral running (see Figure \ref{fig:massFunctionRatio}), and hence dark matter halos should be less biased tracers of the underlying matter density field. This behavior is reversed at low masses for the same reason. Since the effect of a running spectral index on the abundance of structures is stronger at higher redshifts, one might expect the bias ratio to decrease with redshift at high masses and to increase at low masses. What instead is observed is an increase at all masses, however the relation between halo abundance and bias is only a rule-of-thumb, and a direct connection between the mass function ratio and the halo bias ratio cannot be expected, especially at the high mass end where the objects are expected to be extremely rare.

At $z \sim 0$ the halo bias in the model with running of the spectral index can be at most $\sim 5\%$ lower than the bias of the reference model for $M \gtrsim 10^{15} M_\odot h^{-1}$. The linear bias enters quadratically in the correlation function of dark matter halos. Since the observed correlation function of objects at a given redshift is dominated by structures with the lowest mass in the catalogue we can conclude that, e.g.,  a cluster sample containing only very massive objects objects at $z \sim 0$ would have the observed correlation function of galaxy clusters lowered by $\sim 10\%$ if running is turned on. This kind of deviation would be difficult to be detected even with next generation cluster surveys \citep*{FE09.2}. Additionally, uncertainties in the mass-observable relation would blur this signal even more, thus rendering cluster correlations inadequate for model discrimination.

\subsection{Concentration of cosmic structures}\label{sct:res:con}

\begin{figure}
	\includegraphics[width=\hsize]{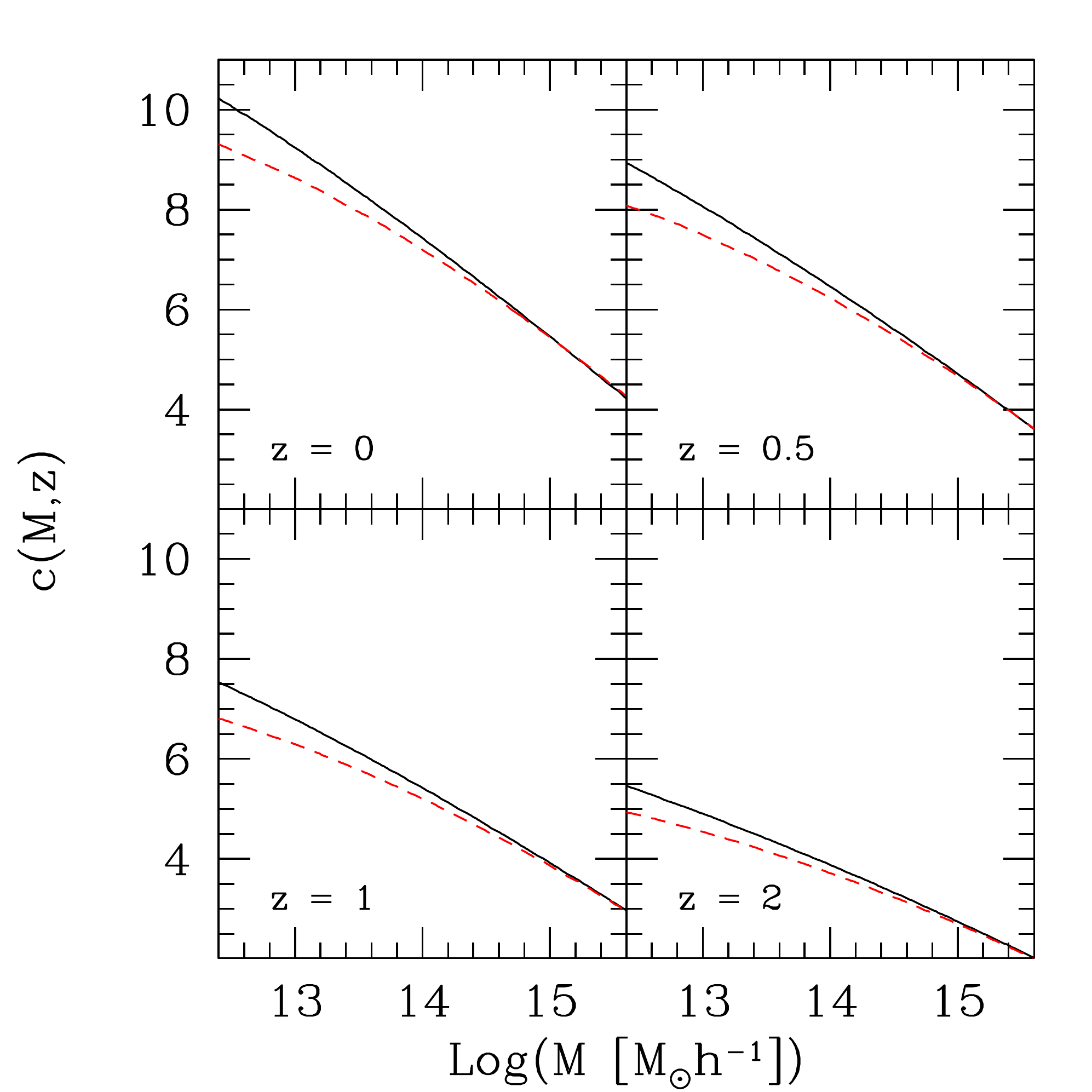}\hfill
	\caption{The concentration of dark matter halos as a function of mass for four different redshifts, as labeled in the plot. The black solid lines show the results for the reference $\Lambda$CDM cosmology, while the red dashed ones refer to the model with running spectral index.}
	\label{fig:haloConcentration}
\end{figure}

In Figure \ref{fig:haloConcentration} we report the relation between mass and concentration of dark matter halos at various redshifts. According to the discussion presented in Section \ref{sct:obs:con} we defined the concentration of an NFW density profile as $c \equiv R_\mathrm{v}/r_\mathrm{s}$, where $R_\mathrm{v}$ is the virial radius and $r_\mathrm{s}$ is the scale radius of the halo. In this plot we adopted the prescription of \citet*{EK01.1} for linking the mass to the concentration, and we did not take into account the extra redshift drop that is necessary to reproduce the power spectrum of numerical simulations (see Section \ref{sct:obs:bao}). 

\begin{figure*}
	\includegraphics[width=0.45\hsize]{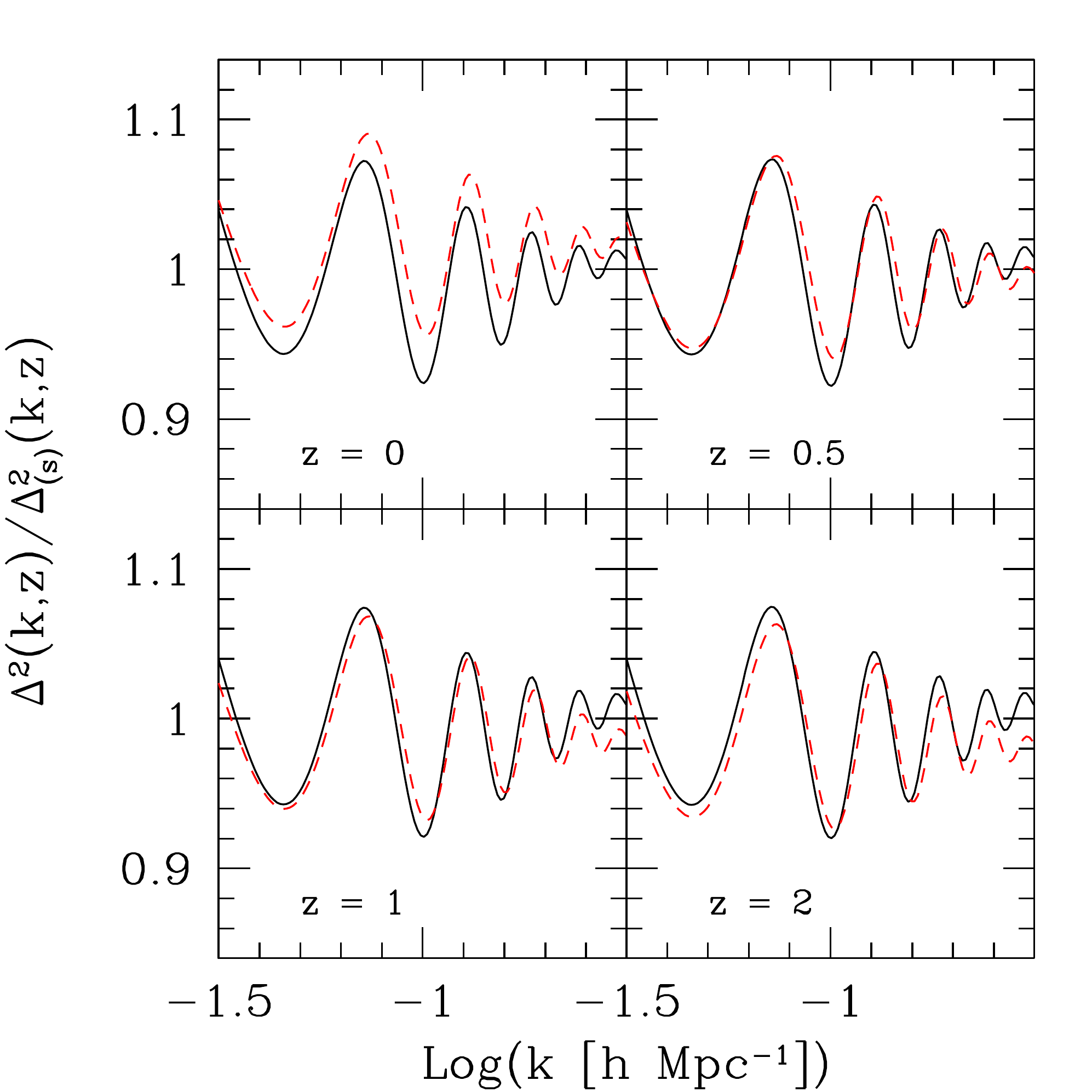}
	\includegraphics[width=0.45\hsize]{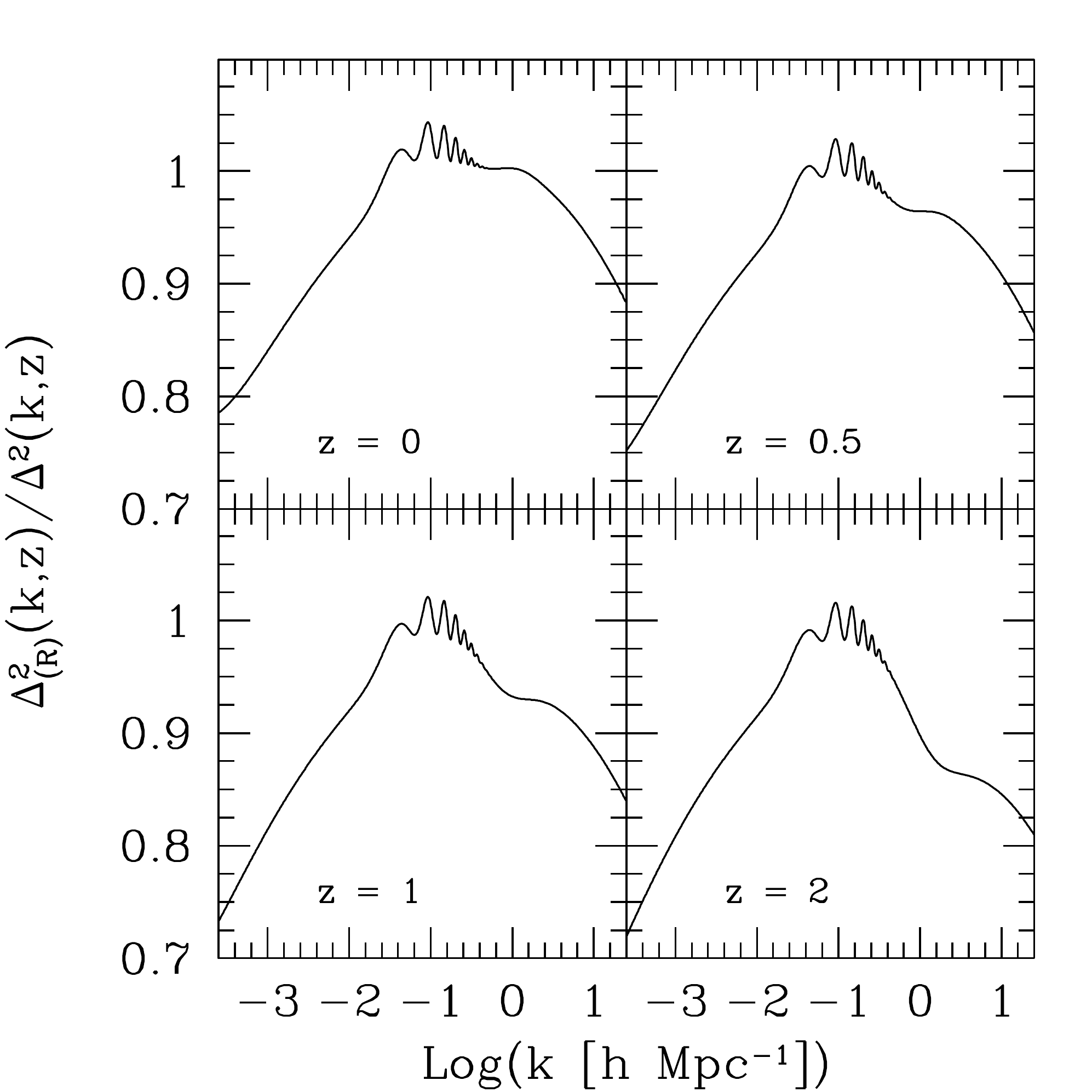}\hfill
	\caption{\emph{Left panel}. The ratio of the fully non-linear dimensionless power in the concordance model (black solid line) and the model with running spectral index (red dashed line) to the smooth power in the concordance model (see text for more details). Results at four different redshifts are shown, as labeled in the plot. \emph{Right panel}. The ratio of the fully non-linear dimensionless powers for the two fiducial cosmological models assumed here.}
	\label{fig:oscillations}
\end{figure*}

As can be seen dark matter halos are generically less concentrated in the model with running of the spectral index when compared to the concordance model. The difference between the two cosmologies
tends to vanish at the high-mass end and it is actually slightly reverted at $z=0$. The drop in concentration can be up to $\sim 10\%$ at galaxy-scale masses. This kind of decrement in dark matter halo concentration can be naively understood in terms of structure formation time. The typical mass $M_*(z)$ of a halo collapsing at a given redshift $z$ is given by $\sigma_{M_*} = \Delta_\mathrm{c}/D_+(z)$. According to Figure \ref{fig:haloConcentration}, at low masses the \emph{rms} of density fluctuations smoothed at a fixed mass scale is lower for the model with running of the spectral index than for the $\Lambda$CDM cosmology. Thus, since $D_+(z)$ is a decreasing function of redshift, the value of $M_*(z)$ for a given $z>0$ is going to be smaller in the former case than in the latter. Turning the argument around, a fixed value of $M_*$ can be reached only at lower redshift in the running model, implying a lower halo concentration. This trend is expected to be reversed at high masses, that is indeed observed in the panel at $z=0$. Of course this is only a rough argument (that besides does not work for $\sigma_M<\Delta_\mathrm{c}$) and it cannot be expected for the extrapolation of a linear scaling relation to account for the complicated physics of non-linear structure formation.

It might be argued that analysis of the concentration-mass relation for galaxies or galaxy clusters could be used to discriminate between the two models at hand. However, given the large errorbars in experimental determinations of the halo concentration and the large intrinsic scatter in the $c-M$ relation \citep{BU07.1,SC07.2}, we believe this is not going to be possible for the moment. Finally, the fact that halo concentrations are lower, implies that the non-linear part of the matter power spectrum should also be suppressed, as we will see in the next Section \ref{sct:res:bao}.

\subsection{Matter power spectrum and Baryon Acoustic Oscillation}\label{sct:res:bao}

We used the procedure detailed in Section \ref{sct:obs:bao} for computing the fully non-linear matter power spectrum. In the left panel of Figure \ref{fig:oscillations} we show the power spectra obtained for the concordance model and the model with running spectral index, relative to the smooth spectrum in the concordance model. This panel is focused in the wavenumber region of the BAO. With smooth spectrum we mean the power spectrum as computed with the transfer function of \cite{SU95.1}, that on the scales of the left panel in Figure \ref{fig:oscillations} is basically identical to the \cite{EI98.1} transfer function, but has no acoustic oscillation. In this way the BAO feature is suitably highlighted. In this Figure and in the remainder of the paper we defined the dimensionless power as

\begin{equation}
\Delta^2(k,z)\equiv \frac{k^3}{2\pi^2}P_\mathrm{NL}(k).
\end{equation}

As can be seen there is a very slight shift, of the order of  a few $\times\:10^{-3} h$ Mpc$^{-1}$, in the position of the BAO, with the acoustic peaks being shifted to somewhat smaller scales due to the presence of spectral running. Also, there is an increment of power at $z=0$ in the running model as compared to the $\Lambda$CDM cosmology, that gradually transforms into a decrement with increasing redshift. The latter effect can be understood as a consequence of the halo model for modeling the non-linear power spectrum. At the scales of the BAO the non-linear effects on power spectrum start to kick in, due to the dark matter particle pairs that are included in the largest halos, while smaller halos only affect larger wavenumbers. As reported in Figure \ref{fig:haloConcentration} the concentration of massive dark matter halos at $z=0$ is slightly larger in the model with running of the primordial spectral index as compared to the concordance model, implying  the increment in power. This trend is reversed at higher redshift since there the concentration of halos in the running model stays always below that in the $\Lambda$CDM cosmology.

\begin{figure*}
	\includegraphics[width=0.45\hsize]{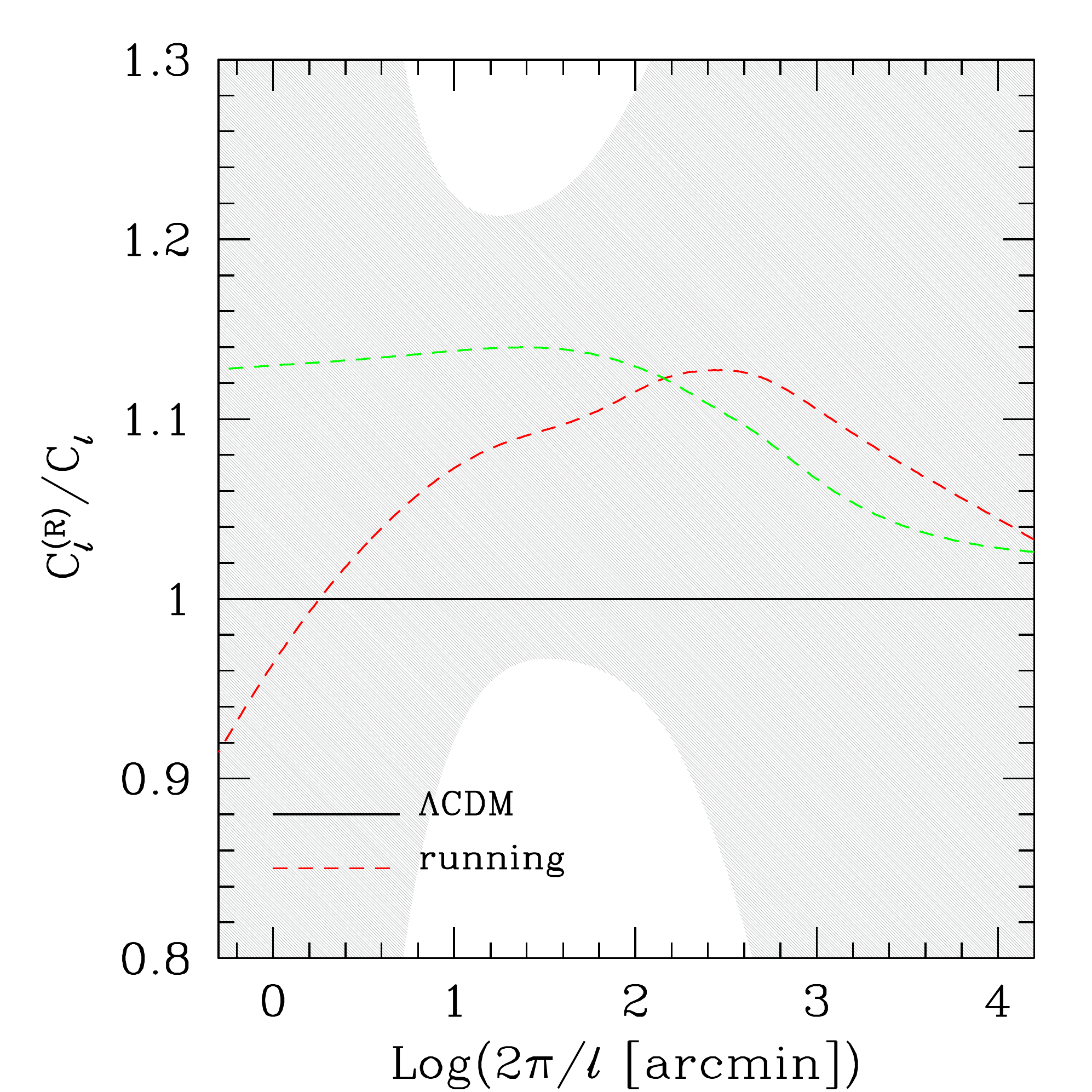}
	\includegraphics[width=0.45\hsize]{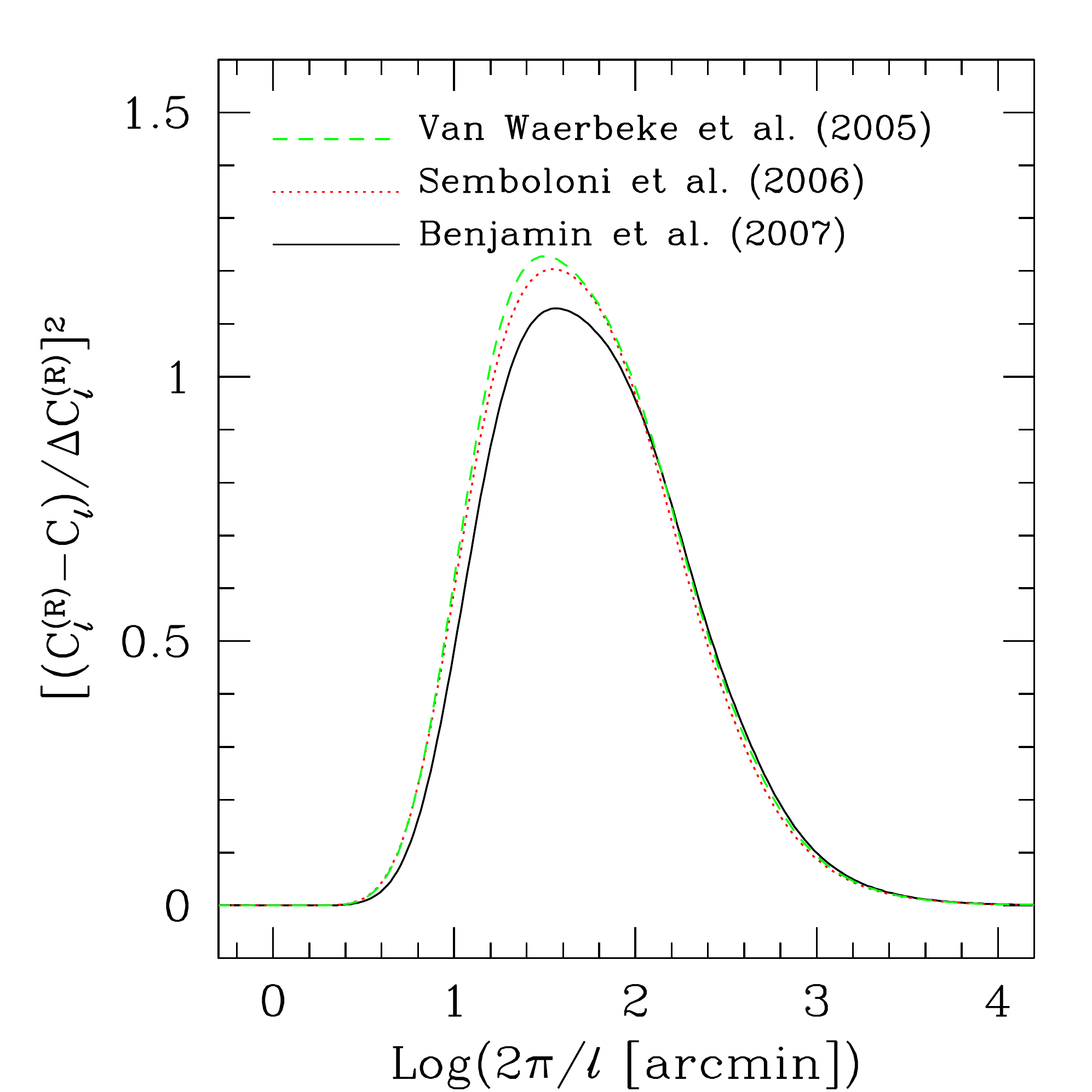}\hfill
	\caption{\emph{Left panel}. The ratio of the weak lensing power spectra computed for the two fiducial cosmological models considered in this work, as labeled in the plot. The gray shaded area represents the statistical error on the ratio obtained via standard error propagation. The green dashed line shows the result of setting $\alpha_{\mathrm{S},0}=0$ in the running model, as discussed in the text. \emph{Right panel}. The signal-to-noise ratio (as defined in the text) for discriminating the cosmological model with running of the primordial spectral index from the standard $\Lambda$CDM cosmology, as a function of multipole.  Results for the three different source redshift distributions detailed in the text are shown, as labeled.}
	\label{fig:weakLensing_R}
\end{figure*}

At larger wavenumbers the power is expected to be suppressed at any redshift, since the effect of smaller mass halos becomes important (refer again to Figure \ref{fig:haloConcentration}) and a suppression is also present in the linear power spectrum, as reported in Figure \ref{fig:matter}. This is indeed shown in the right panel of Figure \ref{fig:oscillations}, where we report the ratio of the non-linear power spectra computed in the two fiducial cosmologies considered here. At the scales of the BAO this ratio is around unity, in agreement with the left panel, and the slight shift in the BAO appears in the wiggles that are present at those scales. The drop in power at very large scales is the same that is observed in the linear power spectrum while, as noted above, the drop at small scales is a combination of the behavior of the linear spectrum and the reduced concentration of dark matter halos.

Errors on the measured matter power spectrum relative to the smooth spectrum in the BAO region are expected to be of the percent level in forthcoming and future galaxy redshift surveys such as BOSS\footnote{\sc http://cosmology.lbl.gov/BOSS/} or EUCLID. Thus, it is unlikely that the position or amplitude of the BAO could be used as a probe of spectral running in the next decade. As for the deviations that are visible at small and large scales in the right panel of Figure \ref{fig:oscillations}, current galaxy redshift surveys cover the wavenumber range $0.01h$ Mpc$^{-1}\lesssim k \lesssim 0.2 h$ Mpc$^{-1}$ \citep{TE04.2,TE06.1}. The power drop that is displayed at small scales remains outside this wavenumber interval, while the low $k$ drop is in principle inside. At such scales however the cosmic variance starts to be significant, and in fact it is larger than the differences we would like to detect. Future surveys might allow to probe scales smaller than the current ones, however robust modeling of the non-linear and baryonic effects on the matter correlations would be required in order to disentangle the effect of a primordial spectral running.

\subsection{Weak lensing power spectrum}\label{sct:res:wl}

We could compute the weak lensing power spectrum according to the prescription detailed in Section \ref{sct:obs:wl}. As noted, this observable depends critically on the adopted source redshift distribution. In the following we assumed the distribution taken from \cite{BE07.2}, unless explicitely noted.

In the left panel of Figure \ref{fig:weakLensing_R} we show the ratio of the weak lensing power spectra obtained for the two models studied in this work, with the statistical error on the ratio evaluated according to standard error propagation. The errors on individuals spectra were computed assuming a EUCLID-like weak lensing survey (see Section \ref{sct:obs:wl}). The ratio between the two spectra is of order $\sim 10\%$ at intermediate scales and drops at both low and high multipoles, resembling the trend in the right panel of Figure \ref{fig:oscillations}. Interestingly, if we perform the change induced by the presence of spectral running on all the cosmological parameters but set the running itself to zero (green dashed line),  we obtain a different behavior, with the power being lifted at small scales and suppressed at large scales with respect to the complete running model. However, we would like to stress again that shifting the cosmological parameters upon introduction of spectral running is self-consistent only if $\alpha_{\mathrm{S},0}$ is indeed left free to be different from zero. Thus, the global effect of running is shown by the red line only. The statistical error is larger than the signal we would like to measure at a fixed multipole. However it should be recalled that in any statistical analysis the weak lensing signal-to-noise ratio is to be summed over hundreds or thousands of multipoles, thus even a signal that is much smaller than that depicted in the left panel of Figure \ref{fig:weakLensing_R} would be significantly detected (see, e.g., \citealt{FE10.1}). This can be better understood in the right panel of the same Figure where we plot the signal-to-noise ratio at a fixed multipole, defined as

\begin{equation}
\frac{S}{N}(\ell)\equiv \left[\frac{C_\ell^\mathrm{(R)}-C_\ell}{\Delta C_\ell}\right]^2,
\end{equation}
for the three different source redshift distributions described in Section \ref{sct:obs:wl}. Obviously, the S/N ratio is of order unity at intermediate scales, hence a signal that is significantly above the noise could be obtained by summing over only a few multipoles. We can therefore conclude that the effect of spectral running could be well detectable with a forthcoming EUCLID-like weak lensing survey.

To end up this section, we would like to stress that the effect of spectral running on the weak lensing power spectrum is still below the effect due to, e.g., variations in the source redshift distribution (see Section \ref{sct:obs:wl}). Hence before being able to discriminate between the two cosmological models with cosmic shear it is necessary to have excellent control on this and other issues, such as the precise prescription to be used for computing the matter power spectrum.

\subsection{Integrated Sachs-Wolfe effect}\label{sct:res:isw}

\begin{figure}
	\includegraphics[width=\hsize]{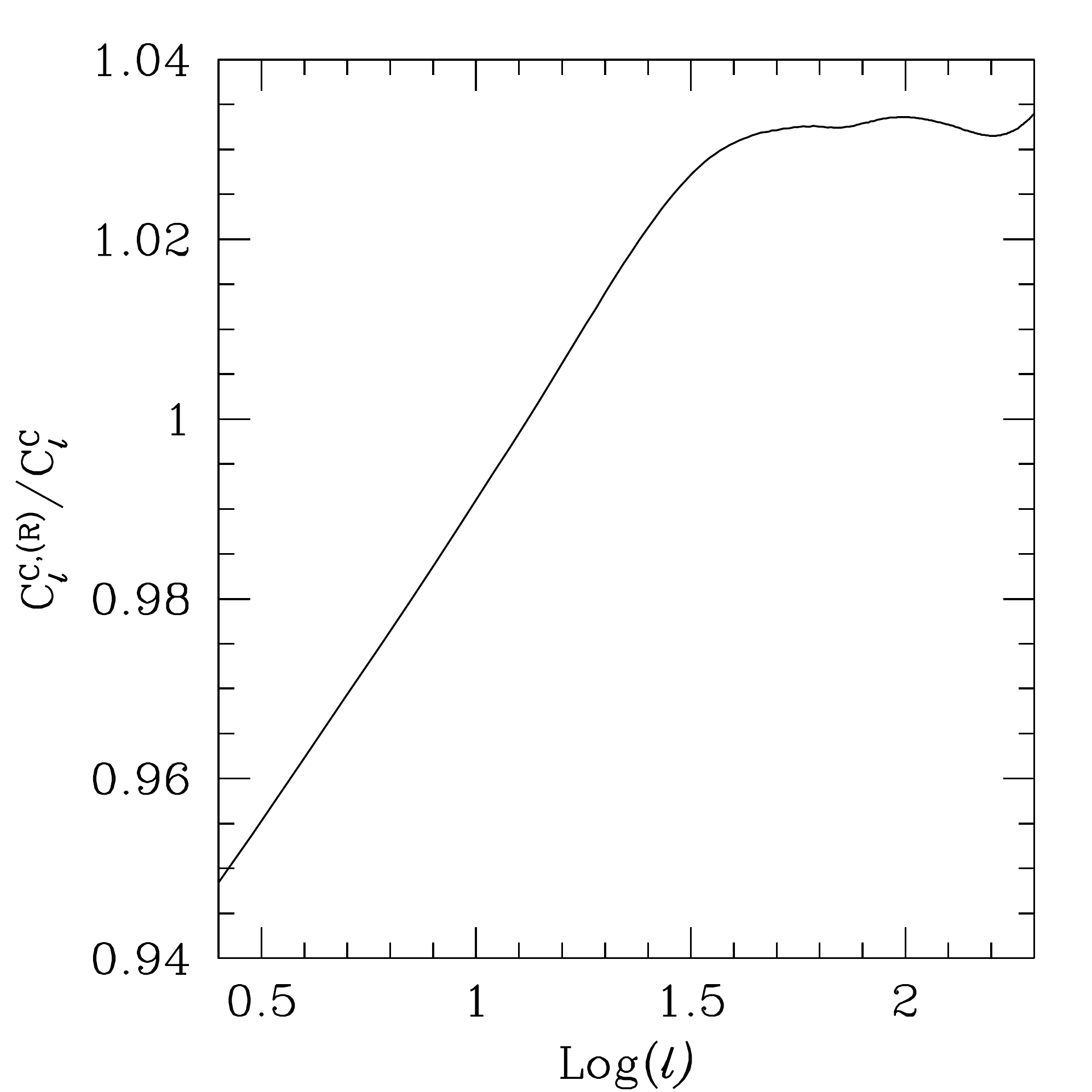}\hfill
	\caption{The ratio of the cross spectrum between the CMB and the LSS due to the ISW effect in the model with running spectral index to the same quantity evaluated in the reference $\Lambda$CDM cosmology.}
	\label{fig:isw}
\end{figure}

In Figure \ref{fig:isw} we report the modifications to the ISW effect that are expected to arise due to primordial spectral running. As can be seen, the cross correlation between the CMB sky and the LSS due to the ISW effect is lower in the running model than the same quantity evaluated in the concordance cosmology at very low multipoles. This trend gets reversed at high $\ell$, however at those scales the absolute amplitude of the ISW effect itself becomes negligible. In any case, the difference between the two models is at most of a few percent at all multipoles.

A few percent deviation is largely smaller than the cosmic variance that dominates the uncertainty in the measured spectrum of the CMB at large scales, as it is apparent from e.g., Figure \ref{fig:Cls}. Therefore we can safely conclude that the effect of primordial spectral running on the ISW effect is not detectable in CMB measurements, and this will remain true in the future as well.

\section{Discussion}\label{sct:alternative}

The choice of the two reference cosmological models that we adopted in our analysis is of course not unique. Rather, the best fit cosmological parameters depend on the datasets that are used, the number 
of parameters that are fitted, and even the running parametrization that is adopted (specifically, the pivot scale $k_\mathrm{p}$). 

For instance, in Section \ref{sct:cosmoparams} it has been shown that adding an additional free parameter to our analysis, the tensor-to-scalar ratio $r$, results in different sets of best-fit cosmological parameters, and in particular in a larger running $\alpha_{\mathrm{S},0}$ of the primordial spectral index. An even larger running can be obtained relaxing the consistency conditions in Eqs. (\ref{eqn:cond1}) and (\ref{eqn:cond2}), i.e. relaxing the assumption of the inflaton as a scalar field with canonical kinetic term. We analyzed the two best-fits in Table \ref{tab:r} and we found that, although the spectral running $\alpha_{\mathrm{S},0}$ is larger than for our fiducial model, the effect thereof on cluster abundance is actually smaller. This result is due to the fact that the process of non-linear structure formation does not depend on the running only, rather it is a complicated interplay of the different cosmological parameters. In particular, according to Table \ref{tab:r} the normalization $\sigma_8$ is slightly larger in the $\Lambda$CDM cosmology than in the model with running of the spectral index, while the opposite is true for the fiducial models we adopted in the rest of this work. As a consequence, the slight increment in $\sigma_M$ at high masses that is displayed in Figure \ref{fig:rmsMass} is not there anymore, and the redshift distributions reported in Figure \ref{fig:clusterAbundance} are basically identical for the two cosmologies.

The cosmological parameters of the best fit reference models depend also on the choice of the pivot scale. It is known that a non-optimal pivot scale might induce unwanted degeneracies in the 
$(n_{\rm S}, \alpha_{{\rm S},0})$ plane \citep*{CO07.1}, which are of course crucial parameters in our analysis. A non-optimal larger pivot scale ($k_\mathrm{p}=0.002 \:{\rm Mpc}^{-1}$) 
has been used by \citet{KO09.1,KO10.1}, obtaining different set of cosmological parameters for the best fits. Similarly to what discussed in the previous case obtained including tensors, $\sigma_8$ happens to be fairly larger in the $\Lambda$CDM case inducing results which are completely different to those presented here. Namely, the abundance of structures in the model with running primordial spectral index is always smaller than in the $\Lambda$CDM cosmology, more so for higher redshift. Indeed, we found that the decrement in the abundance of structures due to spectral running is larger than
the increment found for our fiducial pair of models. This kind of behavior is due to the $\sigma_8$ parameter being substantially smaller in the former case as compared to the latter.

We believe that our analysis reflects anyway the correct and generic qualitative expectations for a small but non-zero running of the scalar spectral index. Our optimal choice of the pivot scale minimizes the shift in $n_{\rm S}$ when the running is added as a cosmological parameter to vary. When running is included we observe a generic increase in the amplitude of scalar perturbations $A_\mathrm{S}$ which in general does not support a substantial decrement in $\sigma_8$, ending in a situation similar to Figure \ref{fig:rmsMass}.

\section{Summary and conclusions}\label{sct:conclusions}

We have explored the statistical properties of non-linear cosmic structures in cosmological models with a spectral index of scalar perturbations which is scale-dependent. We have chosen representative cosmological models - one without and one with running - which are best fits for a combination of CMB and LSS data, i.e. for all the available data which can be probed mainly within the linear gravitational regime. Our main findings can be summarized as follows.

\begin{itemize}
\item The abundance of massive cosmic structures such as galaxy clusters is increased in the model with running spectral index compared to the concordance cosmology, more so for higher redshift and extreme masses. The abundance of rare objects (where "rare" has a specific meaning) can be increased by as much as $\sim 45\%$, and $\sim 30\%$ for objects similar to the massive cluster XMMUJ$2235.3-2557$.
\item Forthcoming galaxy cluster surveys such as the one performed by the X-ray satellite \emph{e}ROSITA and, to a larger extent, the one executed with the submm facility SPT should be able to discriminate between the two models by simple cluster counting. The discriminating power is enhanced if broad redshift information is available, so that high redshift subsamples can be identified within the two catalogues.
\item In agreement with their augmented abundance, cosmic structures are also generically less biased with respect to the underlying matter density field when the running is turned on. This should imply, e.g., a lower amplitude of the correlation function of galaxy clusters, but probably not to the level detectable by future cluster surveys.
\item Dark matter halos are less concentrated in the model with running of the primordial spectral index when compared to the concordance cosmology. The drop in concentration can be up to $\sim 10\%$ on galaxy scales, and is connected with the delayed formation of structures. However, the large errorbars and intrinsic scatter of concentration measurements from, e.g., X-ray emission of galaxy clusters and elliptical galaxies do not allow to draw any meaningful conclusion from this.
\item The lower halo concentration implies a suppression of power in the large scale matter distribution at non-linear scales, which adds up to the suppression present in the linear power spectrum. The BAO presents both a slight shift in the positions of the acoustic peaks and a slight change of amplitude, both however at a level challenging even for future galaxy redshift surveys, such as BOSS and EUCLID.
\item The power spectrum of cosmic shear is increased by $\sim 10\%$ at intermediate angular scales and suppressed at small and large scales, resembling the behavior of the underlying matter power spectrum. Integrating the signal-to-noise ratio around angular separations $\theta \sim 10$ arcmin should bring to a significant detection of the effect of running for a weak lensing survey such as EUCLID. However, excellent control of systematics would be required. 
\item The ISW effect is affected by spectral running only to the percent level, and this is well below the uncertainty due to the cosmic variance that dominates measurements of the CMB power spectrum at very large scales.
\end{itemize}

In conclusion, galaxy cluster counts and the weak lensing power spectrum seem the best options among non-linear observables to constrain a running of the primordial spectral index of density perturbations. We will soon have a drastic improvement in the quality of CMB data by Planck \citep{TH06.1} and of the measurement of $P(k)$ by BOSS and EUCLID which will tighten the constraints on $\alpha_{{\rm S},0}$ based on linear physics. \citet{PA07.2} forecasted that Planck alone will be able to
support the running hypothesis only if $|\alpha_{\mathrm{S},0}| > 0.02$. Our analysis shows that with the combination of present CMB and LSS we are about a factor $1/3-1/2$ with respect to Planck alone capabilities. Our analysis can be transferred in perspective to the near future in which new data on clusters will be available, as it will be for the CMB with Planck. The non-linear observables studied here will be complementary and helpful in the understanding of the initial conditions in the Universe. 

\section*{Acknowledgments}

We acknowledge financial contributions from contracts ASI-INAF I/023/05/0, ASI-INAF I/088/06/0, ASI I/016/07/0 ÒCOFISÓ, and ASI "EUCLID-DUNE" I/064/08/0. We wish to thank F. Schiavon for useful discussions on the impact of running on the ISW effect. We are grateful to the anonymous referee for remarks that allowed us to improve the presentation of our work.

{\small
\bibliographystyle{aa}
\bibliography{./master}
}

\end{document}